\documentclass[longauth]{aaEC}

\usepackage{graphicx}
\usepackage{natbib}         
\usepackage{scalerel}

\usepackage[table]{xcolor}

\usepackage{booktabs}
\usepackage{tabularx}
\usepackage{siunitx}
\usepackage{multirow}
\usepackage{txfonts}
\usepackage[pdfencoding=auto,psdextra]{hyperref}
\hypersetup{
    colorlinks=true,
    linkcolor=blue,
    filecolor=magenta,      
    urlcolor=blue,
    citecolor=blue
}
\usepackage[nameinlink,capitalise]{cleveref}
\crefname{section}{Sect.}{Sects.}
\Crefname{section}{Section}{Sections}
\crefname{figure}{Fig.}{Figs.}
\Crefname{figure}{Figure}{Figures}
\crefname{equation}{Eq.}{Eqs.}
\Crefname{equation}{Equation}{Equations}
\crefname{table}{Table}{Tables}
\crefname{appendix}{Appendix}{Appendices}

\makeatletter
\renewcommand*\aa@pageof{, page \thepage{} of \pageref*{LastPage}}
\makeatother

\usepackage[utf8]{inputenc}

\usepackage[switch, modulo]{lineno}

\usepackage{euclid}

\begin{document}
%
%

\title{{Euclid Quick Data Release (Q1)}}
\subtitle{First study of red quasars selection}



\newcommand{\orcid}[1]{} 
\author{Euclid Collaboration: F.~Tarsitano\orcid{0000-0002-5919-0238}\thanks{\email{federica.tarsitano@unige.ch}}\inst{\ref{aff1}}
\and S.~Fotopoulou\orcid{0000-0002-9686-254X}\inst{\ref{aff2}}
\and M.~Banerji\orcid{0000-0002-0639-5141}\inst{\ref{aff3}}
\and J.~Petley\orcid{0000-0002-4496-0754}\inst{\ref{aff4}}
\and A.~L.~Faisst\orcid{0000-0002-9382-9832}\inst{\ref{aff5}}
\and M.~Tucci\inst{\ref{aff1}}
\and S.~Tacchella\orcid{0000-0002-8224-4505}\inst{\ref{aff6}}
\and Y.~Toba\orcid{0000-0002-3531-7863}\inst{\ref{aff7},\ref{aff8},\ref{aff9}}
\and H.~Landt\orcid{0000-0001-8391-6900}\inst{\ref{aff10}}
\and Y.~Fu\orcid{0000-0002-0759-0504}\inst{\ref{aff4},\ref{aff11}}
\and P.~A.~C.~Cunha\orcid{0000-0002-9454-859X}\inst{\ref{aff12},\ref{aff13}}
\and K.~Duncan\orcid{0000-0001-6889-8388}\inst{\ref{aff14}}
\and W.~Roster\orcid{0000-0002-9149-6528}\inst{\ref{aff15}}
\and M.~Salvato\orcid{0000-0001-7116-9303}\inst{\ref{aff15}}
\and B.~Laloux\orcid{0000-0001-9996-9732}\inst{\ref{aff16},\ref{aff15}}
\and P.~Dayal\orcid{0000-0001-8460-1564}\inst{\ref{aff11}}
\and F.~Ricci\orcid{0000-0001-5742-5980}\inst{\ref{aff17},\ref{aff18}}
\and N.~Aghanim\orcid{0000-0002-6688-8992}\inst{\ref{aff19}}
\and B.~Altieri\orcid{0000-0003-3936-0284}\inst{\ref{aff20}}
\and A.~Amara\inst{\ref{aff21}}
\and S.~Andreon\orcid{0000-0002-2041-8784}\inst{\ref{aff22}}
\and N.~Auricchio\orcid{0000-0003-4444-8651}\inst{\ref{aff23}}
\and H.~Aussel\orcid{0000-0002-1371-5705}\inst{\ref{aff24}}
\and C.~Baccigalupi\orcid{0000-0002-8211-1630}\inst{\ref{aff25},\ref{aff26},\ref{aff27},\ref{aff28}}
\and M.~Baldi\orcid{0000-0003-4145-1943}\inst{\ref{aff29},\ref{aff23},\ref{aff30}}
\and A.~Balestra\orcid{0000-0002-6967-261X}\inst{\ref{aff31}}
\and S.~Bardelli\orcid{0000-0002-8900-0298}\inst{\ref{aff23}}
\and P.~Battaglia\orcid{0000-0002-7337-5909}\inst{\ref{aff23}}
\and A.~Biviano\orcid{0000-0002-0857-0732}\inst{\ref{aff26},\ref{aff25}}
\and A.~Bonchi\orcid{0000-0002-2667-5482}\inst{\ref{aff32}}
\and E.~Branchini\orcid{0000-0002-0808-6908}\inst{\ref{aff33},\ref{aff34},\ref{aff22}}
\and M.~Brescia\orcid{0000-0001-9506-5680}\inst{\ref{aff35},\ref{aff16}}
\and J.~Brinchmann\orcid{0000-0003-4359-8797}\inst{\ref{aff13},\ref{aff12}}
\and S.~Camera\orcid{0000-0003-3399-3574}\inst{\ref{aff36},\ref{aff37},\ref{aff38}}
\and G.~Ca\~nas-Herrera\orcid{0000-0003-2796-2149}\inst{\ref{aff39},\ref{aff40},\ref{aff4}}
\and V.~Capobianco\orcid{0000-0002-3309-7692}\inst{\ref{aff38}}
\and C.~Carbone\orcid{0000-0003-0125-3563}\inst{\ref{aff41}}
\and J.~Carretero\orcid{0000-0002-3130-0204}\inst{\ref{aff42},\ref{aff43}}
\and S.~Casas\orcid{0000-0002-4751-5138}\inst{\ref{aff44}}
\and M.~Castellano\orcid{0000-0001-9875-8263}\inst{\ref{aff18}}
\and G.~Castignani\orcid{0000-0001-6831-0687}\inst{\ref{aff23}}
\and S.~Cavuoti\orcid{0000-0002-3787-4196}\inst{\ref{aff16},\ref{aff45}}
\and K.~C.~Chambers\orcid{0000-0001-6965-7789}\inst{\ref{aff46}}
\and A.~Cimatti\inst{\ref{aff47}}
\and C.~Colodro-Conde\inst{\ref{aff48}}
\and G.~Congedo\orcid{0000-0003-2508-0046}\inst{\ref{aff14}}
\and C.~J.~Conselice\orcid{0000-0003-1949-7638}\inst{\ref{aff49}}
\and L.~Conversi\orcid{0000-0002-6710-8476}\inst{\ref{aff50},\ref{aff20}}
\and Y.~Copin\orcid{0000-0002-5317-7518}\inst{\ref{aff51}}
\and A.~Costille\inst{\ref{aff52}}
\and F.~Courbin\orcid{0000-0003-0758-6510}\inst{\ref{aff53},\ref{aff54}}
\and H.~M.~Courtois\orcid{0000-0003-0509-1776}\inst{\ref{aff55}}
\and M.~Cropper\orcid{0000-0003-4571-9468}\inst{\ref{aff56}}
\and A.~Da~Silva\orcid{0000-0002-6385-1609}\inst{\ref{aff57},\ref{aff58}}
\and H.~Degaudenzi\orcid{0000-0002-5887-6799}\inst{\ref{aff1}}
\and G.~De~Lucia\orcid{0000-0002-6220-9104}\inst{\ref{aff26}}
\and A.~M.~Di~Giorgio\orcid{0000-0002-4767-2360}\inst{\ref{aff59}}
\and C.~Dolding\orcid{0009-0003-7199-6108}\inst{\ref{aff56}}
\and H.~Dole\orcid{0000-0002-9767-3839}\inst{\ref{aff19}}
\and F.~Dubath\orcid{0000-0002-6533-2810}\inst{\ref{aff1}}
\and C.~A.~J.~Duncan\orcid{0009-0003-3573-0791}\inst{\ref{aff49}}
\and X.~Dupac\inst{\ref{aff20}}
\and S.~Escoffier\orcid{0000-0002-2847-7498}\inst{\ref{aff60}}
\and M.~Fabricius\orcid{0000-0002-7025-6058}\inst{\ref{aff15},\ref{aff61}}
\and M.~Farina\orcid{0000-0002-3089-7846}\inst{\ref{aff59}}
\and R.~Farinelli\inst{\ref{aff23}}
\and F.~Faustini\orcid{0000-0001-6274-5145}\inst{\ref{aff32},\ref{aff18}}
\and S.~Ferriol\inst{\ref{aff51}}
\and F.~Finelli\orcid{0000-0002-6694-3269}\inst{\ref{aff23},\ref{aff62}}
\and M.~Frailis\orcid{0000-0002-7400-2135}\inst{\ref{aff26}}
\and E.~Franceschi\orcid{0000-0002-0585-6591}\inst{\ref{aff23}}
\and S.~Galeotta\orcid{0000-0002-3748-5115}\inst{\ref{aff26}}
\and K.~George\orcid{0000-0002-1734-8455}\inst{\ref{aff61}}
\and W.~Gillard\orcid{0000-0003-4744-9748}\inst{\ref{aff60}}
\and B.~Gillis\orcid{0000-0002-4478-1270}\inst{\ref{aff14}}
\and C.~Giocoli\orcid{0000-0002-9590-7961}\inst{\ref{aff23},\ref{aff30}}
\and P.~G\'omez-Alvarez\orcid{0000-0002-8594-5358}\inst{\ref{aff63},\ref{aff20}}
\and J.~Gracia-Carpio\inst{\ref{aff15}}
\and B.~R.~Granett\orcid{0000-0003-2694-9284}\inst{\ref{aff22}}
\and A.~Grazian\orcid{0000-0002-5688-0663}\inst{\ref{aff31}}
\and F.~Grupp\inst{\ref{aff15},\ref{aff61}}
\and L.~Guzzo\orcid{0000-0001-8264-5192}\inst{\ref{aff64},\ref{aff22},\ref{aff65}}
\and S.~Gwyn\orcid{0000-0001-8221-8406}\inst{\ref{aff66}}
\and S.~V.~H.~Haugan\orcid{0000-0001-9648-7260}\inst{\ref{aff67}}
\and W.~Holmes\inst{\ref{aff68}}
\and I.~M.~Hook\orcid{0000-0002-2960-978X}\inst{\ref{aff69}}
\and F.~Hormuth\inst{\ref{aff70}}
\and A.~Hornstrup\orcid{0000-0002-3363-0936}\inst{\ref{aff71},\ref{aff72}}
\and P.~Hudelot\inst{\ref{aff73}}
\and K.~Jahnke\orcid{0000-0003-3804-2137}\inst{\ref{aff74}}
\and M.~Jhabvala\inst{\ref{aff75}}
\and E.~Keih\"anen\orcid{0000-0003-1804-7715}\inst{\ref{aff76}}
\and S.~Kermiche\orcid{0000-0002-0302-5735}\inst{\ref{aff60}}
\and A.~Kiessling\orcid{0000-0002-2590-1273}\inst{\ref{aff68}}
\and B.~Kubik\orcid{0009-0006-5823-4880}\inst{\ref{aff51}}
\and M.~K\"ummel\orcid{0000-0003-2791-2117}\inst{\ref{aff61}}
\and M.~Kunz\orcid{0000-0002-3052-7394}\inst{\ref{aff77}}
\and H.~Kurki-Suonio\orcid{0000-0002-4618-3063}\inst{\ref{aff78},\ref{aff79}}
\and Q.~Le~Boulc'h\inst{\ref{aff80}}
\and A.~M.~C.~Le~Brun\orcid{0000-0002-0936-4594}\inst{\ref{aff81}}
\and D.~Le~Mignant\orcid{0000-0002-5339-5515}\inst{\ref{aff52}}
\and S.~Ligori\orcid{0000-0003-4172-4606}\inst{\ref{aff38}}
\and P.~B.~Lilje\orcid{0000-0003-4324-7794}\inst{\ref{aff67}}
\and V.~Lindholm\orcid{0000-0003-2317-5471}\inst{\ref{aff78},\ref{aff79}}
\and I.~Lloro\orcid{0000-0001-5966-1434}\inst{\ref{aff82}}
\and G.~Mainetti\orcid{0000-0003-2384-2377}\inst{\ref{aff80}}
\and D.~Maino\inst{\ref{aff64},\ref{aff41},\ref{aff65}}
\and E.~Maiorano\orcid{0000-0003-2593-4355}\inst{\ref{aff23}}
\and O.~Mansutti\orcid{0000-0001-5758-4658}\inst{\ref{aff26}}
\and S.~Marcin\inst{\ref{aff83}}
\and O.~Marggraf\orcid{0000-0001-7242-3852}\inst{\ref{aff84}}
\and M.~Martinelli\orcid{0000-0002-6943-7732}\inst{\ref{aff18},\ref{aff85}}
\and N.~Martinet\orcid{0000-0003-2786-7790}\inst{\ref{aff52}}
\and F.~Marulli\orcid{0000-0002-8850-0303}\inst{\ref{aff86},\ref{aff23},\ref{aff30}}
\and R.~Massey\orcid{0000-0002-6085-3780}\inst{\ref{aff87}}
\and E.~Medinaceli\orcid{0000-0002-4040-7783}\inst{\ref{aff23}}
\and S.~Mei\orcid{0000-0002-2849-559X}\inst{\ref{aff88},\ref{aff89}}
\and M.~Melchior\inst{\ref{aff90}}
\and Y.~Mellier\inst{\ref{aff91},\ref{aff73}}
\and M.~Meneghetti\orcid{0000-0003-1225-7084}\inst{\ref{aff23},\ref{aff30}}
\and E.~Merlin\orcid{0000-0001-6870-8900}\inst{\ref{aff18}}
\and G.~Meylan\inst{\ref{aff92}}
\and A.~Mora\orcid{0000-0002-1922-8529}\inst{\ref{aff93}}
\and M.~Moresco\orcid{0000-0002-7616-7136}\inst{\ref{aff86},\ref{aff23}}
\and L.~Moscardini\orcid{0000-0002-3473-6716}\inst{\ref{aff86},\ref{aff23},\ref{aff30}}
\and R.~Nakajima\orcid{0009-0009-1213-7040}\inst{\ref{aff84}}
\and C.~Neissner\orcid{0000-0001-8524-4968}\inst{\ref{aff94},\ref{aff43}}
\and S.-M.~Niemi\inst{\ref{aff39}}
\and J.~W.~Nightingale\orcid{0000-0002-8987-7401}\inst{\ref{aff95}}
\and C.~Padilla\orcid{0000-0001-7951-0166}\inst{\ref{aff94}}
\and S.~Paltani\orcid{0000-0002-8108-9179}\inst{\ref{aff1}}
\and F.~Pasian\orcid{0000-0002-4869-3227}\inst{\ref{aff26}}
\and K.~Pedersen\inst{\ref{aff96}}
\and W.~J.~Percival\orcid{0000-0002-0644-5727}\inst{\ref{aff97},\ref{aff98},\ref{aff99}}
\and V.~Pettorino\inst{\ref{aff39}}
\and S.~Pires\orcid{0000-0002-0249-2104}\inst{\ref{aff24}}
\and G.~Polenta\orcid{0000-0003-4067-9196}\inst{\ref{aff32}}
\and M.~Poncet\inst{\ref{aff100}}
\and L.~A.~Popa\inst{\ref{aff101}}
\and L.~Pozzetti\orcid{0000-0001-7085-0412}\inst{\ref{aff23}}
\and F.~Raison\orcid{0000-0002-7819-6918}\inst{\ref{aff15}}
\and R.~Rebolo\orcid{0000-0003-3767-7085}\inst{\ref{aff48},\ref{aff102},\ref{aff103}}
\and A.~Renzi\orcid{0000-0001-9856-1970}\inst{\ref{aff104},\ref{aff105}}
\and J.~Rhodes\orcid{0000-0002-4485-8549}\inst{\ref{aff68}}
\and G.~Riccio\inst{\ref{aff16}}
\and E.~Romelli\orcid{0000-0003-3069-9222}\inst{\ref{aff26}}
\and M.~Roncarelli\orcid{0000-0001-9587-7822}\inst{\ref{aff23}}
\and E.~Rossetti\orcid{0000-0003-0238-4047}\inst{\ref{aff29}}
\and B.~Rusholme\orcid{0000-0001-7648-4142}\inst{\ref{aff5}}
\and R.~Saglia\orcid{0000-0003-0378-7032}\inst{\ref{aff61},\ref{aff15}}
\and Z.~Sakr\orcid{0000-0002-4823-3757}\inst{\ref{aff106},\ref{aff107},\ref{aff108}}
\and D.~Sapone\orcid{0000-0001-7089-4503}\inst{\ref{aff109}}
\and B.~Sartoris\orcid{0000-0003-1337-5269}\inst{\ref{aff61},\ref{aff26}}
\and J.~A.~Schewtschenko\orcid{0000-0002-4913-6393}\inst{\ref{aff14}}
\and P.~Schneider\orcid{0000-0001-8561-2679}\inst{\ref{aff84}}
\and T.~Schrabback\orcid{0000-0002-6987-7834}\inst{\ref{aff110}}
\and M.~Scodeggio\inst{\ref{aff41}}
\and A.~Secroun\orcid{0000-0003-0505-3710}\inst{\ref{aff60}}
\and G.~Seidel\orcid{0000-0003-2907-353X}\inst{\ref{aff74}}
\and M.~Seiffert\orcid{0000-0002-7536-9393}\inst{\ref{aff68}}
\and S.~Serrano\orcid{0000-0002-0211-2861}\inst{\ref{aff111},\ref{aff112},\ref{aff113}}
\and P.~Simon\inst{\ref{aff84}}
\and C.~Sirignano\orcid{0000-0002-0995-7146}\inst{\ref{aff104},\ref{aff105}}
\and G.~Sirri\orcid{0000-0003-2626-2853}\inst{\ref{aff30}}
\and A.~Spurio~Mancini\orcid{0000-0001-5698-0990}\inst{\ref{aff114}}
\and L.~Stanco\orcid{0000-0002-9706-5104}\inst{\ref{aff105}}
\and J.~Steinwagner\orcid{0000-0001-7443-1047}\inst{\ref{aff15}}
\and P.~Tallada-Cresp\'{i}\orcid{0000-0002-1336-8328}\inst{\ref{aff42},\ref{aff43}}
\and A.~N.~Taylor\inst{\ref{aff14}}
\and H.~I.~Teplitz\orcid{0000-0002-7064-5424}\inst{\ref{aff115}}
\and I.~Tereno\inst{\ref{aff57},\ref{aff116}}
\and S.~Toft\orcid{0000-0003-3631-7176}\inst{\ref{aff117},\ref{aff118}}
\and R.~Toledo-Moreo\orcid{0000-0002-2997-4859}\inst{\ref{aff119}}
\and F.~Torradeflot\orcid{0000-0003-1160-1517}\inst{\ref{aff43},\ref{aff42}}
\and I.~Tutusaus\orcid{0000-0002-3199-0399}\inst{\ref{aff107}}
\and L.~Valenziano\orcid{0000-0002-1170-0104}\inst{\ref{aff23},\ref{aff62}}
\and J.~Valiviita\orcid{0000-0001-6225-3693}\inst{\ref{aff78},\ref{aff79}}
\and T.~Vassallo\orcid{0000-0001-6512-6358}\inst{\ref{aff61},\ref{aff26}}
\and G.~Verdoes~Kleijn\orcid{0000-0001-5803-2580}\inst{\ref{aff11}}
\and A.~Veropalumbo\orcid{0000-0003-2387-1194}\inst{\ref{aff22},\ref{aff34},\ref{aff33}}
\and Y.~Wang\orcid{0000-0002-4749-2984}\inst{\ref{aff115}}
\and J.~Weller\orcid{0000-0002-8282-2010}\inst{\ref{aff61},\ref{aff15}}
\and A.~Zacchei\orcid{0000-0003-0396-1192}\inst{\ref{aff26},\ref{aff25}}
\and G.~Zamorani\orcid{0000-0002-2318-301X}\inst{\ref{aff23}}
\and F.~M.~Zerbi\inst{\ref{aff22}}
\and E.~Zucca\orcid{0000-0002-5845-8132}\inst{\ref{aff23}}
\and V.~Allevato\orcid{0000-0001-7232-5152}\inst{\ref{aff16}}
\and M.~Ballardini\orcid{0000-0003-4481-3559}\inst{\ref{aff120},\ref{aff121},\ref{aff23}}
\and M.~Bolzonella\orcid{0000-0003-3278-4607}\inst{\ref{aff23}}
\and E.~Bozzo\orcid{0000-0002-8201-1525}\inst{\ref{aff1}}
\and C.~Burigana\orcid{0000-0002-3005-5796}\inst{\ref{aff122},\ref{aff62}}
\and R.~Cabanac\orcid{0000-0001-6679-2600}\inst{\ref{aff107}}
\and A.~Cappi\inst{\ref{aff23},\ref{aff123}}
\and D.~Di~Ferdinando\inst{\ref{aff30}}
\and J.~A.~Escartin~Vigo\inst{\ref{aff15}}
\and L.~Gabarra\orcid{0000-0002-8486-8856}\inst{\ref{aff124}}
\and J.~Mart\'{i}n-Fleitas\orcid{0000-0002-8594-569X}\inst{\ref{aff93}}
\and S.~Matthew\orcid{0000-0001-8448-1697}\inst{\ref{aff14}}
\and N.~Mauri\orcid{0000-0001-8196-1548}\inst{\ref{aff47},\ref{aff30}}
\and R.~B.~Metcalf\orcid{0000-0003-3167-2574}\inst{\ref{aff86},\ref{aff23}}
\and A.~Pezzotta\orcid{0000-0003-0726-2268}\inst{\ref{aff125},\ref{aff15}}
\and M.~P\"ontinen\orcid{0000-0001-5442-2530}\inst{\ref{aff78}}
\and C.~Porciani\orcid{0000-0002-7797-2508}\inst{\ref{aff84}}
\and I.~Risso\orcid{0000-0003-2525-7761}\inst{\ref{aff126}}
\and V.~Scottez\inst{\ref{aff91},\ref{aff127}}
\and M.~Sereno\orcid{0000-0003-0302-0325}\inst{\ref{aff23},\ref{aff30}}
\and M.~Tenti\orcid{0000-0002-4254-5901}\inst{\ref{aff30}}
\and M.~Viel\orcid{0000-0002-2642-5707}\inst{\ref{aff25},\ref{aff26},\ref{aff28},\ref{aff27},\ref{aff128}}
\and M.~Wiesmann\orcid{0009-0000-8199-5860}\inst{\ref{aff67}}
\and Y.~Akrami\orcid{0000-0002-2407-7956}\inst{\ref{aff129},\ref{aff130}}
\and I.~T.~Andika\orcid{0000-0001-6102-9526}\inst{\ref{aff131},\ref{aff132}}
\and S.~Anselmi\orcid{0000-0002-3579-9583}\inst{\ref{aff105},\ref{aff104},\ref{aff133}}
\and M.~Archidiacono\orcid{0000-0003-4952-9012}\inst{\ref{aff64},\ref{aff65}}
\and F.~Atrio-Barandela\orcid{0000-0002-2130-2513}\inst{\ref{aff134}}
\and C.~Benoist\inst{\ref{aff123}}
\and K.~Benson\inst{\ref{aff56}}
\and D.~Bertacca\orcid{0000-0002-2490-7139}\inst{\ref{aff104},\ref{aff31},\ref{aff105}}
\and M.~Bethermin\orcid{0000-0002-3915-2015}\inst{\ref{aff135}}
\and L.~Bisigello\orcid{0000-0003-0492-4924}\inst{\ref{aff31}}
\and A.~Blanchard\orcid{0000-0001-8555-9003}\inst{\ref{aff107}}
\and L.~Blot\orcid{0000-0002-9622-7167}\inst{\ref{aff136},\ref{aff133}}
\and M.~L.~Brown\orcid{0000-0002-0370-8077}\inst{\ref{aff49}}
\and S.~Bruton\orcid{0000-0002-6503-5218}\inst{\ref{aff137}}
\and A.~Calabro\orcid{0000-0003-2536-1614}\inst{\ref{aff18}}
\and F.~Caro\inst{\ref{aff18}}
\and C.~S.~Carvalho\inst{\ref{aff116}}
\and T.~Castro\orcid{0000-0002-6292-3228}\inst{\ref{aff26},\ref{aff27},\ref{aff25},\ref{aff128}}
\and Y.~Charles\inst{\ref{aff52}}
\and F.~Cogato\orcid{0000-0003-4632-6113}\inst{\ref{aff86},\ref{aff23}}
\and A.~R.~Cooray\orcid{0000-0002-3892-0190}\inst{\ref{aff138}}
\and O.~Cucciati\orcid{0000-0002-9336-7551}\inst{\ref{aff23}}
\and S.~Davini\orcid{0000-0003-3269-1718}\inst{\ref{aff34}}
\and F.~De~Paolis\orcid{0000-0001-6460-7563}\inst{\ref{aff139},\ref{aff140},\ref{aff141}}
\and G.~Desprez\orcid{0000-0001-8325-1742}\inst{\ref{aff11}}
\and A.~D\'iaz-S\'anchez\orcid{0000-0003-0748-4768}\inst{\ref{aff142}}
\and J.~J.~Diaz\inst{\ref{aff143}}
\and S.~Di~Domizio\orcid{0000-0003-2863-5895}\inst{\ref{aff33},\ref{aff34}}
\and J.~M.~Diego\orcid{0000-0001-9065-3926}\inst{\ref{aff144}}
\and P.-A.~Duc\orcid{0000-0003-3343-6284}\inst{\ref{aff135}}
\and A.~Enia\orcid{0000-0002-0200-2857}\inst{\ref{aff29},\ref{aff23}}
\and Y.~Fang\inst{\ref{aff61}}
\and A.~G.~Ferrari\orcid{0009-0005-5266-4110}\inst{\ref{aff30}}
\and A.~Finoguenov\orcid{0000-0002-4606-5403}\inst{\ref{aff78}}
\and A.~Fontana\orcid{0000-0003-3820-2823}\inst{\ref{aff18}}
\and A.~Franco\orcid{0000-0002-4761-366X}\inst{\ref{aff140},\ref{aff139},\ref{aff141}}
\and K.~Ganga\orcid{0000-0001-8159-8208}\inst{\ref{aff88}}
\and J.~Garc\'ia-Bellido\orcid{0000-0002-9370-8360}\inst{\ref{aff129}}
\and T.~Gasparetto\orcid{0000-0002-7913-4866}\inst{\ref{aff26}}
\and V.~Gautard\inst{\ref{aff145}}
\and E.~Gaztanaga\orcid{0000-0001-9632-0815}\inst{\ref{aff113},\ref{aff111},\ref{aff146}}
\and F.~Giacomini\orcid{0000-0002-3129-2814}\inst{\ref{aff30}}
\and F.~Gianotti\orcid{0000-0003-4666-119X}\inst{\ref{aff23}}
\and G.~Gozaliasl\orcid{0000-0002-0236-919X}\inst{\ref{aff147},\ref{aff78}}
\and M.~Guidi\orcid{0000-0001-9408-1101}\inst{\ref{aff29},\ref{aff23}}
\and C.~M.~Gutierrez\orcid{0000-0001-7854-783X}\inst{\ref{aff148}}
\and A.~Hall\orcid{0000-0002-3139-8651}\inst{\ref{aff14}}
\and C.~Hern\'andez-Monteagudo\orcid{0000-0001-5471-9166}\inst{\ref{aff103},\ref{aff48}}
\and H.~Hildebrandt\orcid{0000-0002-9814-3338}\inst{\ref{aff149}}
\and J.~Hjorth\orcid{0000-0002-4571-2306}\inst{\ref{aff96}}
\and J.~J.~E.~Kajava\orcid{0000-0002-3010-8333}\inst{\ref{aff150},\ref{aff151}}
\and Y.~Kang\orcid{0009-0000-8588-7250}\inst{\ref{aff1}}
\and V.~Kansal\orcid{0000-0002-4008-6078}\inst{\ref{aff152},\ref{aff153}}
\and D.~Karagiannis\orcid{0000-0002-4927-0816}\inst{\ref{aff120},\ref{aff154}}
\and K.~Kiiveri\inst{\ref{aff76}}
\and C.~C.~Kirkpatrick\inst{\ref{aff76}}
\and S.~Kruk\orcid{0000-0001-8010-8879}\inst{\ref{aff20}}
\and J.~Le~Graet\orcid{0000-0001-6523-7971}\inst{\ref{aff60}}
\and L.~Legrand\orcid{0000-0003-0610-5252}\inst{\ref{aff155},\ref{aff6}}
\and M.~Lembo\orcid{0000-0002-5271-5070}\inst{\ref{aff120},\ref{aff121}}
\and F.~Lepori\orcid{0009-0000-5061-7138}\inst{\ref{aff156}}
\and G.~Leroy\orcid{0009-0004-2523-4425}\inst{\ref{aff10},\ref{aff87}}
\and G.~F.~Lesci\orcid{0000-0002-4607-2830}\inst{\ref{aff86},\ref{aff23}}
\and J.~Lesgourgues\orcid{0000-0001-7627-353X}\inst{\ref{aff44}}
\and L.~Leuzzi\orcid{0009-0006-4479-7017}\inst{\ref{aff86},\ref{aff23}}
\and T.~I.~Liaudat\orcid{0000-0002-9104-314X}\inst{\ref{aff157}}
\and S.~J.~Liu\orcid{0000-0001-7680-2139}\inst{\ref{aff59}}
\and A.~Loureiro\orcid{0000-0002-4371-0876}\inst{\ref{aff158},\ref{aff159}}
\and J.~Macias-Perez\orcid{0000-0002-5385-2763}\inst{\ref{aff160}}
\and G.~Maggio\orcid{0000-0003-4020-4836}\inst{\ref{aff26}}
\and M.~Magliocchetti\orcid{0000-0001-9158-4838}\inst{\ref{aff59}}
\and F.~Mannucci\orcid{0000-0002-4803-2381}\inst{\ref{aff161}}
\and R.~Maoli\orcid{0000-0002-6065-3025}\inst{\ref{aff162},\ref{aff18}}
\and C.~J.~A.~P.~Martins\orcid{0000-0002-4886-9261}\inst{\ref{aff163},\ref{aff13}}
\and L.~Maurin\orcid{0000-0002-8406-0857}\inst{\ref{aff19}}
\and M.~Miluzio\inst{\ref{aff20},\ref{aff164}}
\and P.~Monaco\orcid{0000-0003-2083-7564}\inst{\ref{aff165},\ref{aff26},\ref{aff27},\ref{aff25}}
\and C.~Moretti\orcid{0000-0003-3314-8936}\inst{\ref{aff28},\ref{aff128},\ref{aff26},\ref{aff25},\ref{aff27}}
\and G.~Morgante\inst{\ref{aff23}}
\and S.~Nadathur\orcid{0000-0001-9070-3102}\inst{\ref{aff146}}
\and K.~Naidoo\orcid{0000-0002-9182-1802}\inst{\ref{aff146}}
\and A.~Navarro-Alsina\orcid{0000-0002-3173-2592}\inst{\ref{aff84}}
\and S.~Nesseris\orcid{0000-0002-0567-0324}\inst{\ref{aff129}}
\and F.~Passalacqua\orcid{0000-0002-8606-4093}\inst{\ref{aff104},\ref{aff105}}
\and K.~Paterson\orcid{0000-0001-8340-3486}\inst{\ref{aff74}}
\and L.~Patrizii\inst{\ref{aff30}}
\and A.~Pisani\orcid{0000-0002-6146-4437}\inst{\ref{aff60},\ref{aff166}}
\and D.~Potter\orcid{0000-0002-0757-5195}\inst{\ref{aff156}}
\and S.~Quai\orcid{0000-0002-0449-8163}\inst{\ref{aff86},\ref{aff23}}
\and M.~Radovich\orcid{0000-0002-3585-866X}\inst{\ref{aff31}}
\and P.-F.~Rocci\inst{\ref{aff19}}
\and G.~Rodighiero\orcid{0000-0002-9415-2296}\inst{\ref{aff104},\ref{aff31}}
\and S.~Sacquegna\orcid{0000-0002-8433-6630}\inst{\ref{aff139},\ref{aff140},\ref{aff141}}
\and M.~Sahl\'en\orcid{0000-0003-0973-4804}\inst{\ref{aff167}}
\and D.~B.~Sanders\orcid{0000-0002-1233-9998}\inst{\ref{aff46}}
\and E.~Sarpa\orcid{0000-0002-1256-655X}\inst{\ref{aff28},\ref{aff128},\ref{aff27}}
\and A.~Schneider\orcid{0000-0001-7055-8104}\inst{\ref{aff156}}
\and M.~Schultheis\inst{\ref{aff123}}
\and D.~Sciotti\orcid{0009-0008-4519-2620}\inst{\ref{aff18},\ref{aff85}}
\and E.~Sellentin\inst{\ref{aff168},\ref{aff4}}
\and F.~Shankar\orcid{0000-0001-8973-5051}\inst{\ref{aff3}}
\and L.~C.~Smith\orcid{0000-0002-3259-2771}\inst{\ref{aff169}}
\and K.~Tanidis\orcid{0000-0001-9843-5130}\inst{\ref{aff124}}
\and G.~Testera\inst{\ref{aff34}}
\and R.~Teyssier\orcid{0000-0001-7689-0933}\inst{\ref{aff166}}
\and S.~Tosi\orcid{0000-0002-7275-9193}\inst{\ref{aff33},\ref{aff126}}
\and A.~Troja\orcid{0000-0003-0239-4595}\inst{\ref{aff104},\ref{aff105}}
\and C.~Valieri\inst{\ref{aff30}}
\and A.~Venhola\orcid{0000-0001-6071-4564}\inst{\ref{aff170}}
\and D.~Vergani\orcid{0000-0003-0898-2216}\inst{\ref{aff23}}
\and G.~Verza\orcid{0000-0002-1886-8348}\inst{\ref{aff171}}
\and P.~Vielzeuf\orcid{0000-0003-2035-9339}\inst{\ref{aff60}}
\and A.~Viitanen\orcid{0000-0001-9383-786X}\inst{\ref{aff76},\ref{aff18}}
\and N.~A.~Walton\orcid{0000-0003-3983-8778}\inst{\ref{aff169}}
\and J.~G.~Sorce\orcid{0000-0002-2307-2432}\inst{\ref{aff172},\ref{aff19}}}
										   
\institute{Department of Astronomy, University of Geneva, ch. d'Ecogia 16, 1290 Versoix, Switzerland\label{aff1}
\and
School of Physics, HH Wills Physics Laboratory, University of Bristol, Tyndall Avenue, Bristol, BS8 1TL, UK\label{aff2}
\and
School of Physics \& Astronomy, University of Southampton, Highfield Campus, Southampton SO17 1BJ, UK\label{aff3}
\and
Leiden Observatory, Leiden University, Einsteinweg 55, 2333 CC Leiden, The Netherlands\label{aff4}
\and
Caltech/IPAC, 1200 E. California Blvd., Pasadena, CA 91125, USA\label{aff5}
\and
Kavli Institute for Cosmology Cambridge, Madingley Road, Cambridge, CB3 0HA, UK\label{aff6}
\and
Department of Physical Sciences, Ritsumeikan University, Kusatsu, Shiga 525-8577, Japan\label{aff7}
\and
National Astronomical Observatory of Japan, 2-21-1 Osawa, Mitaka, Tokyo 181-8588, Japan\label{aff8}
\and
Academia Sinica Institute of Astronomy and Astrophysics (ASIAA), 11F of ASMAB, No.~1, Section 4, Roosevelt Road, Taipei 10617, Taiwan\label{aff9}
\and
Department of Physics, Centre for Extragalactic Astronomy, Durham University, South Road, Durham, DH1 3LE, UK\label{aff10}
\and
Kapteyn Astronomical Institute, University of Groningen, PO Box 800, 9700 AV Groningen, The Netherlands\label{aff11}
\and
Faculdade de Ci\^encias da Universidade do Porto, Rua do Campo de Alegre, 4150-007 Porto, Portugal\label{aff12}
\and
Instituto de Astrof\'isica e Ci\^encias do Espa\c{c}o, Universidade do Porto, CAUP, Rua das Estrelas, PT4150-762 Porto, Portugal\label{aff13}
\and
Institute for Astronomy, University of Edinburgh, Royal Observatory, Blackford Hill, Edinburgh EH9 3HJ, UK\label{aff14}
\and
Max Planck Institute for Extraterrestrial Physics, Giessenbachstr. 1, 85748 Garching, Germany\label{aff15}
\and
INAF-Osservatorio Astronomico di Capodimonte, Via Moiariello 16, 80131 Napoli, Italy\label{aff16}
\and
Department of Mathematics and Physics, Roma Tre University, Via della Vasca Navale 84, 00146 Rome, Italy\label{aff17}
\and
INAF-Osservatorio Astronomico di Roma, Via Frascati 33, 00078 Monteporzio Catone, Italy\label{aff18}
\and
Universit\'e Paris-Saclay, CNRS, Institut d'astrophysique spatiale, 91405, Orsay, France\label{aff19}
\and
ESAC/ESA, Camino Bajo del Castillo, s/n., Urb. Villafranca del Castillo, 28692 Villanueva de la Ca\~nada, Madrid, Spain\label{aff20}
\and
School of Mathematics and Physics, University of Surrey, Guildford, Surrey, GU2 7XH, UK\label{aff21}
\and
INAF-Osservatorio Astronomico di Brera, Via Brera 28, 20122 Milano, Italy\label{aff22}
\and
INAF-Osservatorio di Astrofisica e Scienza dello Spazio di Bologna, Via Piero Gobetti 93/3, 40129 Bologna, Italy\label{aff23}
\and
Universit\'e Paris-Saclay, Universit\'e Paris Cit\'e, CEA, CNRS, AIM, 91191, Gif-sur-Yvette, France\label{aff24}
\and
IFPU, Institute for Fundamental Physics of the Universe, via Beirut 2, 34151 Trieste, Italy\label{aff25}
\and
INAF-Osservatorio Astronomico di Trieste, Via G. B. Tiepolo 11, 34143 Trieste, Italy\label{aff26}
\and
INFN, Sezione di Trieste, Via Valerio 2, 34127 Trieste TS, Italy\label{aff27}
\and
SISSA, International School for Advanced Studies, Via Bonomea 265, 34136 Trieste TS, Italy\label{aff28}
\and
Dipartimento di Fisica e Astronomia, Universit\`a di Bologna, Via Gobetti 93/2, 40129 Bologna, Italy\label{aff29}
\and
INFN-Sezione di Bologna, Viale Berti Pichat 6/2, 40127 Bologna, Italy\label{aff30}
\and
INAF-Osservatorio Astronomico di Padova, Via dell'Osservatorio 5, 35122 Padova, Italy\label{aff31}
\and
Space Science Data Center, Italian Space Agency, via del Politecnico snc, 00133 Roma, Italy\label{aff32}
\and
Dipartimento di Fisica, Universit\`a di Genova, Via Dodecaneso 33, 16146, Genova, Italy\label{aff33}
\and
INFN-Sezione di Genova, Via Dodecaneso 33, 16146, Genova, Italy\label{aff34}
\and
Department of Physics "E. Pancini", University Federico II, Via Cinthia 6, 80126, Napoli, Italy\label{aff35}
\and
Dipartimento di Fisica, Universit\`a degli Studi di Torino, Via P. Giuria 1, 10125 Torino, Italy\label{aff36}
\and
INFN-Sezione di Torino, Via P. Giuria 1, 10125 Torino, Italy\label{aff37}
\and
INAF-Osservatorio Astrofisico di Torino, Via Osservatorio 20, 10025 Pino Torinese (TO), Italy\label{aff38}
\and
European Space Agency/ESTEC, Keplerlaan 1, 2201 AZ Noordwijk, The Netherlands\label{aff39}
\and
Institute Lorentz, Leiden University, Niels Bohrweg 2, 2333 CA Leiden, The Netherlands\label{aff40}
\and
INAF-IASF Milano, Via Alfonso Corti 12, 20133 Milano, Italy\label{aff41}
\and
Centro de Investigaciones Energ\'eticas, Medioambientales y Tecnol\'ogicas (CIEMAT), Avenida Complutense 40, 28040 Madrid, Spain\label{aff42}
\and
Port d'Informaci\'{o} Cient\'{i}fica, Campus UAB, C. Albareda s/n, 08193 Bellaterra (Barcelona), Spain\label{aff43}
\and
Institute for Theoretical Particle Physics and Cosmology (TTK), RWTH Aachen University, 52056 Aachen, Germany\label{aff44}
\and
INFN section of Naples, Via Cinthia 6, 80126, Napoli, Italy\label{aff45}
\and
Institute for Astronomy, University of Hawaii, 2680 Woodlawn Drive, Honolulu, HI 96822, USA\label{aff46}
\and
Dipartimento di Fisica e Astronomia "Augusto Righi" - Alma Mater Studiorum Universit\`a di Bologna, Viale Berti Pichat 6/2, 40127 Bologna, Italy\label{aff47}
\and
Instituto de Astrof\'{\i}sica de Canarias, V\'{\i}a L\'actea, 38205 La Laguna, Tenerife, Spain\label{aff48}
\and
Jodrell Bank Centre for Astrophysics, Department of Physics and Astronomy, University of Manchester, Oxford Road, Manchester M13 9PL, UK\label{aff49}
\and
European Space Agency/ESRIN, Largo Galileo Galilei 1, 00044 Frascati, Roma, Italy\label{aff50}
\and
Universit\'e Claude Bernard Lyon 1, CNRS/IN2P3, IP2I Lyon, UMR 5822, Villeurbanne, F-69100, France\label{aff51}
\and
Aix-Marseille Universit\'e, CNRS, CNES, LAM, Marseille, France\label{aff52}
\and
Institut de Ci\`{e}ncies del Cosmos (ICCUB), Universitat de Barcelona (IEEC-UB), Mart\'{i} i Franqu\`{e}s 1, 08028 Barcelona, Spain\label{aff53}
\and
Instituci\'o Catalana de Recerca i Estudis Avan\c{c}ats (ICREA), Passeig de Llu\'{\i}s Companys 23, 08010 Barcelona, Spain\label{aff54}
\and
UCB Lyon 1, CNRS/IN2P3, IUF, IP2I Lyon, 4 rue Enrico Fermi, 69622 Villeurbanne, France\label{aff55}
\and
Mullard Space Science Laboratory, University College London, Holmbury St Mary, Dorking, Surrey RH5 6NT, UK\label{aff56}
\and
Departamento de F\'isica, Faculdade de Ci\^encias, Universidade de Lisboa, Edif\'icio C8, Campo Grande, PT1749-016 Lisboa, Portugal\label{aff57}
\and
Instituto de Astrof\'isica e Ci\^encias do Espa\c{c}o, Faculdade de Ci\^encias, Universidade de Lisboa, Campo Grande, 1749-016 Lisboa, Portugal\label{aff58}
\and
INAF-Istituto di Astrofisica e Planetologia Spaziali, via del Fosso del Cavaliere, 100, 00100 Roma, Italy\label{aff59}
\and
Aix-Marseille Universit\'e, CNRS/IN2P3, CPPM, Marseille, France\label{aff60}
\and
Universit\"ats-Sternwarte M\"unchen, Fakult\"at f\"ur Physik, Ludwig-Maximilians-Universit\"at M\"unchen, Scheinerstrasse 1, 81679 M\"unchen, Germany\label{aff61}
\and
INFN-Bologna, Via Irnerio 46, 40126 Bologna, Italy\label{aff62}
\and
FRACTAL S.L.N.E., calle Tulip\'an 2, Portal 13 1A, 28231, Las Rozas de Madrid, Spain\label{aff63}
\and
Dipartimento di Fisica "Aldo Pontremoli", Universit\`a degli Studi di Milano, Via Celoria 16, 20133 Milano, Italy\label{aff64}
\and
INFN-Sezione di Milano, Via Celoria 16, 20133 Milano, Italy\label{aff65}
\and
NRC Herzberg, 5071 West Saanich Rd, Victoria, BC V9E 2E7, Canada\label{aff66}
\and
Institute of Theoretical Astrophysics, University of Oslo, P.O. Box 1029 Blindern, 0315 Oslo, Norway\label{aff67}
\and
Jet Propulsion Laboratory, California Institute of Technology, 4800 Oak Grove Drive, Pasadena, CA, 91109, USA\label{aff68}
\and
Department of Physics, Lancaster University, Lancaster, LA1 4YB, UK\label{aff69}
\and
Felix Hormuth Engineering, Goethestr. 17, 69181 Leimen, Germany\label{aff70}
\and
Technical University of Denmark, Elektrovej 327, 2800 Kgs. Lyngby, Denmark\label{aff71}
\and
Cosmic Dawn Center (DAWN), Denmark\label{aff72}
\and
Institut d'Astrophysique de Paris, UMR 7095, CNRS, and Sorbonne Universit\'e, 98 bis boulevard Arago, 75014 Paris, France\label{aff73}
\and
Max-Planck-Institut f\"ur Astronomie, K\"onigstuhl 17, 69117 Heidelberg, Germany\label{aff74}
\and
NASA Goddard Space Flight Center, Greenbelt, MD 20771, USA\label{aff75}
\and
Department of Physics and Helsinki Institute of Physics, Gustaf H\"allstr\"omin katu 2, 00014 University of Helsinki, Finland\label{aff76}
\and
Universit\'e de Gen\`eve, D\'epartement de Physique Th\'eorique and Centre for Astroparticle Physics, 24 quai Ernest-Ansermet, CH-1211 Gen\`eve 4, Switzerland\label{aff77}
\and
Department of Physics, P.O. Box 64, 00014 University of Helsinki, Finland\label{aff78}
\and
Helsinki Institute of Physics, Gustaf H{\"a}llstr{\"o}min katu 2, University of Helsinki, Helsinki, Finland\label{aff79}
\and
Centre de Calcul de l'IN2P3/CNRS, 21 avenue Pierre de Coubertin 69627 Villeurbanne Cedex, France\label{aff80}
\and
Laboratoire d'etude de l'Univers et des phenomenes eXtremes, Observatoire de Paris, Universit\'e PSL, Sorbonne Universit\'e, CNRS, 92190 Meudon, France\label{aff81}
\and
SKA Observatory, Jodrell Bank, Lower Withington, Macclesfield, Cheshire SK11 9FT, UK\label{aff82}
\and
University of Applied Sciences and Arts of Northwestern Switzerland, School of Computer Science, 5210 Windisch, Switzerland\label{aff83}
\and
Universit\"at Bonn, Argelander-Institut f\"ur Astronomie, Auf dem H\"ugel 71, 53121 Bonn, Germany\label{aff84}
\and
INFN-Sezione di Roma, Piazzale Aldo Moro, 2 - c/o Dipartimento di Fisica, Edificio G. Marconi, 00185 Roma, Italy\label{aff85}
\and
Dipartimento di Fisica e Astronomia "Augusto Righi" - Alma Mater Studiorum Universit\`a di Bologna, via Piero Gobetti 93/2, 40129 Bologna, Italy\label{aff86}
\and
Department of Physics, Institute for Computational Cosmology, Durham University, South Road, Durham, DH1 3LE, UK\label{aff87}
\and
Universit\'e Paris Cit\'e, CNRS, Astroparticule et Cosmologie, 75013 Paris, France\label{aff88}
\and
CNRS-UCB International Research Laboratory, Centre Pierre Binetruy, IRL2007, CPB-IN2P3, Berkeley, USA\label{aff89}
\and
University of Applied Sciences and Arts of Northwestern Switzerland, School of Engineering, 5210 Windisch, Switzerland\label{aff90}
\and
Institut d'Astrophysique de Paris, 98bis Boulevard Arago, 75014, Paris, France\label{aff91}
\and
Institute of Physics, Laboratory of Astrophysics, Ecole Polytechnique F\'ed\'erale de Lausanne (EPFL), Observatoire de Sauverny, 1290 Versoix, Switzerland\label{aff92}
\and
Aurora Technology for European Space Agency (ESA), Camino bajo del Castillo, s/n, Urbanizacion Villafranca del Castillo, Villanueva de la Ca\~nada, 28692 Madrid, Spain\label{aff93}
\and
Institut de F\'{i}sica d'Altes Energies (IFAE), The Barcelona Institute of Science and Technology, Campus UAB, 08193 Bellaterra (Barcelona), Spain\label{aff94}
\and
School of Mathematics, Statistics and Physics, Newcastle University, Herschel Building, Newcastle-upon-Tyne, NE1 7RU, UK\label{aff95}
\and
DARK, Niels Bohr Institute, University of Copenhagen, Jagtvej 155, 2200 Copenhagen, Denmark\label{aff96}
\and
Waterloo Centre for Astrophysics, University of Waterloo, Waterloo, Ontario N2L 3G1, Canada\label{aff97}
\and
Department of Physics and Astronomy, University of Waterloo, Waterloo, Ontario N2L 3G1, Canada\label{aff98}
\and
Perimeter Institute for Theoretical Physics, Waterloo, Ontario N2L 2Y5, Canada\label{aff99}
\and
Centre National d'Etudes Spatiales -- Centre spatial de Toulouse, 18 avenue Edouard Belin, 31401 Toulouse Cedex 9, France\label{aff100}
\and
Institute of Space Science, Str. Atomistilor, nr. 409 M\u{a}gurele, Ilfov, 077125, Romania\label{aff101}
\and
Consejo Superior de Investigaciones Cientificas, Calle Serrano 117, 28006 Madrid, Spain\label{aff102}
\and
Universidad de La Laguna, Departamento de Astrof\'{\i}sica, 38206 La Laguna, Tenerife, Spain\label{aff103}
\and
Dipartimento di Fisica e Astronomia "G. Galilei", Universit\`a di Padova, Via Marzolo 8, 35131 Padova, Italy\label{aff104}
\and
INFN-Padova, Via Marzolo 8, 35131 Padova, Italy\label{aff105}
\and
Institut f\"ur Theoretische Physik, University of Heidelberg, Philosophenweg 16, 69120 Heidelberg, Germany\label{aff106}
\and
Institut de Recherche en Astrophysique et Plan\'etologie (IRAP), Universit\'e de Toulouse, CNRS, UPS, CNES, 14 Av. Edouard Belin, 31400 Toulouse, France\label{aff107}
\and
Universit\'e St Joseph; Faculty of Sciences, Beirut, Lebanon\label{aff108}
\and
Departamento de F\'isica, FCFM, Universidad de Chile, Blanco Encalada 2008, Santiago, Chile\label{aff109}
\and
Universit\"at Innsbruck, Institut f\"ur Astro- und Teilchenphysik, Technikerstr. 25/8, 6020 Innsbruck, Austria\label{aff110}
\and
Institut d'Estudis Espacials de Catalunya (IEEC),  Edifici RDIT, Campus UPC, 08860 Castelldefels, Barcelona, Spain\label{aff111}
\and
Satlantis, University Science Park, Sede Bld 48940, Leioa-Bilbao, Spain\label{aff112}
\and
Institute of Space Sciences (ICE, CSIC), Campus UAB, Carrer de Can Magrans, s/n, 08193 Barcelona, Spain\label{aff113}
\and
Department of Physics, Royal Holloway, University of London, TW20 0EX, UK\label{aff114}
\and
Infrared Processing and Analysis Center, California Institute of Technology, Pasadena, CA 91125, USA\label{aff115}
\and
Instituto de Astrof\'isica e Ci\^encias do Espa\c{c}o, Faculdade de Ci\^encias, Universidade de Lisboa, Tapada da Ajuda, 1349-018 Lisboa, Portugal\label{aff116}
\and
Cosmic Dawn Center (DAWN)\label{aff117}
\and
Niels Bohr Institute, University of Copenhagen, Jagtvej 128, 2200 Copenhagen, Denmark\label{aff118}
\and
Universidad Polit\'ecnica de Cartagena, Departamento de Electr\'onica y Tecnolog\'ia de Computadoras,  Plaza del Hospital 1, 30202 Cartagena, Spain\label{aff119}
\and
Dipartimento di Fisica e Scienze della Terra, Universit\`a degli Studi di Ferrara, Via Giuseppe Saragat 1, 44122 Ferrara, Italy\label{aff120}
\and
Istituto Nazionale di Fisica Nucleare, Sezione di Ferrara, Via Giuseppe Saragat 1, 44122 Ferrara, Italy\label{aff121}
\and
INAF, Istituto di Radioastronomia, Via Piero Gobetti 101, 40129 Bologna, Italy\label{aff122}
\and
Universit\'e C\^{o}te d'Azur, Observatoire de la C\^{o}te d'Azur, CNRS, Laboratoire Lagrange, Bd de l'Observatoire, CS 34229, 06304 Nice cedex 4, France\label{aff123}
\and
Department of Physics, Oxford University, Keble Road, Oxford OX1 3RH, UK\label{aff124}
\and
INAF - Osservatorio Astronomico di Brera, via Emilio Bianchi 46, 23807 Merate, Italy\label{aff125}
\and
INAF-Osservatorio Astronomico di Brera, Via Brera 28, 20122 Milano, Italy, and INFN-Sezione di Genova, Via Dodecaneso 33, 16146, Genova, Italy\label{aff126}
\and
ICL, Junia, Universit\'e Catholique de Lille, LITL, 59000 Lille, France\label{aff127}
\and
ICSC - Centro Nazionale di Ricerca in High Performance Computing, Big Data e Quantum Computing, Via Magnanelli 2, Bologna, Italy\label{aff128}
\and
Instituto de F\'isica Te\'orica UAM-CSIC, Campus de Cantoblanco, 28049 Madrid, Spain\label{aff129}
\and
CERCA/ISO, Department of Physics, Case Western Reserve University, 10900 Euclid Avenue, Cleveland, OH 44106, USA\label{aff130}
\and
Technical University of Munich, TUM School of Natural Sciences, Physics Department, James-Franck-Str.~1, 85748 Garching, Germany\label{aff131}
\and
Max-Planck-Institut f\"ur Astrophysik, Karl-Schwarzschild-Str.~1, 85748 Garching, Germany\label{aff132}
\and
Laboratoire Univers et Th\'eorie, Observatoire de Paris, Universit\'e PSL, Universit\'e Paris Cit\'e, CNRS, 92190 Meudon, France\label{aff133}
\and
Departamento de F{\'\i}sica Fundamental. Universidad de Salamanca. Plaza de la Merced s/n. 37008 Salamanca, Spain\label{aff134}
\and
Universit\'e de Strasbourg, CNRS, Observatoire astronomique de Strasbourg, UMR 7550, 67000 Strasbourg, France\label{aff135}
\and
Center for Data-Driven Discovery, Kavli IPMU (WPI), UTIAS, The University of Tokyo, Kashiwa, Chiba 277-8583, Japan\label{aff136}
\and
California Institute of Technology, 1200 E California Blvd, Pasadena, CA 91125, USA\label{aff137}
\and
Department of Physics \& Astronomy, University of California Irvine, Irvine CA 92697, USA\label{aff138}
\and
Department of Mathematics and Physics E. De Giorgi, University of Salento, Via per Arnesano, CP-I93, 73100, Lecce, Italy\label{aff139}
\and
INFN, Sezione di Lecce, Via per Arnesano, CP-193, 73100, Lecce, Italy\label{aff140}
\and
INAF-Sezione di Lecce, c/o Dipartimento Matematica e Fisica, Via per Arnesano, 73100, Lecce, Italy\label{aff141}
\and
Departamento F\'isica Aplicada, Universidad Polit\'ecnica de Cartagena, Campus Muralla del Mar, 30202 Cartagena, Murcia, Spain\label{aff142}
\and
Instituto de Astrof\'isica de Canarias (IAC); Departamento de Astrof\'isica, Universidad de La Laguna (ULL), 38200, La Laguna, Tenerife, Spain\label{aff143}
\and
Instituto de F\'isica de Cantabria, Edificio Juan Jord\'a, Avenida de los Castros, 39005 Santander, Spain\label{aff144}
\and
CEA Saclay, DFR/IRFU, Service d'Astrophysique, Bat. 709, 91191 Gif-sur-Yvette, France\label{aff145}
\and
Institute of Cosmology and Gravitation, University of Portsmouth, Portsmouth PO1 3FX, UK\label{aff146}
\and
Department of Computer Science, Aalto University, PO Box 15400, Espoo, FI-00 076, Finland\label{aff147}
\and
Instituto de Astrof\'\i sica de Canarias, c/ Via Lactea s/n, La Laguna 38200, Spain. Departamento de Astrof\'\i sica de la Universidad de La Laguna, Avda. Francisco Sanchez, La Laguna, 38200, Spain\label{aff148}
\and
Ruhr University Bochum, Faculty of Physics and Astronomy, Astronomical Institute (AIRUB), German Centre for Cosmological Lensing (GCCL), 44780 Bochum, Germany\label{aff149}
\and
Department of Physics and Astronomy, Vesilinnantie 5, 20014 University of Turku, Finland\label{aff150}
\and
Serco for European Space Agency (ESA), Camino bajo del Castillo, s/n, Urbanizacion Villafranca del Castillo, Villanueva de la Ca\~nada, 28692 Madrid, Spain\label{aff151}
\and
ARC Centre of Excellence for Dark Matter Particle Physics, Melbourne, Australia\label{aff152}
\and
Centre for Astrophysics \& Supercomputing, Swinburne University of Technology,  Hawthorn, Victoria 3122, Australia\label{aff153}
\and
Department of Physics and Astronomy, University of the Western Cape, Bellville, Cape Town, 7535, South Africa\label{aff154}
\and
DAMTP, Centre for Mathematical Sciences, Wilberforce Road, Cambridge CB3 0WA, UK\label{aff155}
\and
Department of Astrophysics, University of Zurich, Winterthurerstrasse 190, 8057 Zurich, Switzerland\label{aff156}
\and
IRFU, CEA, Universit\'e Paris-Saclay 91191 Gif-sur-Yvette Cedex, France\label{aff157}
\and
Oskar Klein Centre for Cosmoparticle Physics, Department of Physics, Stockholm University, Stockholm, SE-106 91, Sweden\label{aff158}
\and
Astrophysics Group, Blackett Laboratory, Imperial College London, London SW7 2AZ, UK\label{aff159}
\and
Univ. Grenoble Alpes, CNRS, Grenoble INP, LPSC-IN2P3, 53, Avenue des Martyrs, 38000, Grenoble, France\label{aff160}
\and
INAF-Osservatorio Astrofisico di Arcetri, Largo E. Fermi 5, 50125, Firenze, Italy\label{aff161}
\and
Dipartimento di Fisica, Sapienza Universit\`a di Roma, Piazzale Aldo Moro 2, 00185 Roma, Italy\label{aff162}
\and
Centro de Astrof\'{\i}sica da Universidade do Porto, Rua das Estrelas, 4150-762 Porto, Portugal\label{aff163}
\and
HE Space for European Space Agency (ESA), Camino bajo del Castillo, s/n, Urbanizacion Villafranca del Castillo, Villanueva de la Ca\~nada, 28692 Madrid, Spain\label{aff164}
\and
Dipartimento di Fisica - Sezione di Astronomia, Universit\`a di Trieste, Via Tiepolo 11, 34131 Trieste, Italy\label{aff165}
\and
Department of Astrophysical Sciences, Peyton Hall, Princeton University, Princeton, NJ 08544, USA\label{aff166}
\and
Theoretical astrophysics, Department of Physics and Astronomy, Uppsala University, Box 515, 751 20 Uppsala, Sweden\label{aff167}
\and
Mathematical Institute, University of Leiden, Einsteinweg 55, 2333 CA Leiden, The Netherlands\label{aff168}
\and
Institute of Astronomy, University of Cambridge, Madingley Road, Cambridge CB3 0HA, UK\label{aff169}
\and
Space physics and astronomy research unit, University of Oulu, Pentti Kaiteran katu 1, FI-90014 Oulu, Finland\label{aff170}
\and
Center for Computational Astrophysics, Flatiron Institute, 162 5th Avenue, 10010, New York, NY, USA\label{aff171}
\and
Univ. Lille, CNRS, Centrale Lille, UMR 9189 CRIStAL, 59000 Lille, France\label{aff172}}        

%
%
 \abstract{

Red quasars constitute an important but elusive phase in the evolution of supermassive black holes, where dust obscuration can significantly alter their observed properties. They have broad emission lines, like other quasars, but their optical continuum emission is significantly reddened, which is why they were traditionally identified based on near- and mid-infrared selection criteria.
This work showcases the capability of the \Euclid space telescope to find a large sample of red quasars, using \Euclid near infrared (NIR) photometry. We first conduct a forecast analysis, comparing a synthetic catalogue of red QSOs with COSMOS2020. Using template fitting, we reconstruct \Euclid-like photometry for the COSMOS sources and identify a sample of candidates in a multidimensional colour-colour space achieving $98\%$ completeness for mock red QSOs with $30\%$ contaminants. To refine our selection function, we implement a probabilistic Random Forest classifier, and use UMAP visualisation to disentangle non-linear features in colour-space, reaching $98\%$ completeness and $88\%$ purity. A preliminary analysis of the candidates in the \Euclid Deep Field Fornax (EDF-F) shows that, compared to VISTA+DECAm-based colour selection criteria, \Euclid’s superior depth, resolution and optical-to-NIR coverage improves the identification of the reddest, most obscured sources. Notably, the \Euclid exquisite resolution in the $I_E$ filter unveils the presence of a candidate dual quasar system, highlighting the potential for this mission to contribute to future studies on the population of dual AGN.
The resulting catalogue of candidates, including more the 150 000 sources, provides a first census of red quasars in \Euclid Q1 and sets the groundwork for future studies in the Euclid Wide Survey (EWS), including spectral follow-up analyses and host morphology characterisation. 


}
%
%
\keywords{Galaxies: evolution, active; quasars: general, supermassive black holes; Methods: statistical, numerical.}
%
%
   \titlerunning{\Euclid Q1 - Red quasars}
   \authorrunning{Euclid Collaboration: F. Tarsitano et al.}
   
   \maketitle
%
%
%
%
   
\section{\label{sc:Intro}Introduction}
Supermassive black holes ($M_{\rm BH} > 10^{6} \  M_{\odot}$) and their host galaxies are believed to grow in tandem, as postulated by theoretical \citep[e.g.,][]{1998A&A...331L...1S} and observational \citep[e.g.,][]{1998AJ....115.2285M} arguments. At the same time, early galaxy evolution simulations showed that in order to reproduce observed galaxy size and brightness distributions, as well as their star formation, some form of energetic feedback from the central black hole is required \citep{2006MNRAS.370..645B, 2006MNRAS.369.1808C}. Black holes can influence their host galaxies through the release of gravitational potential energy in the form of radiation during their active phase of accretion of matter, also known as active galactic nuclei (AGN). In addition, some AGN are known to power energetic jets and winds which extend their influence on their host galaxy. Therefore, AGN hold the missing piece to advance our knowledge of the black hole-galaxy co-evolution \citep{2012ARA&A..50..455F, 2014ARA&A..52..589H}.

The lack of a detailed theory of AGN feedback leads to unrealistic models, failing to capture the observed complexity of the AGN population. For example, models do not reliably predict the ratio of unobscured (type 1; face-on view of the accretion disc) to obscured (type 2; edge-on view of the accretion disc and obscuring torus) AGN, nor the evolution of their luminosity function. \citet{2021MNRAS.503.1940H} presented six current state-of-the-art galaxy evolution simulations compared to the number density of AGN derived from X-ray observations. Even though these models are successful in predicting the observed properties of normal galaxies, none of them predicts reliably the history of supermassive black hole growth and the corresponding AGN phase. Thus, the creation of an AGN activity model rooted in observations is needed to act as ground truth for galaxy evolution simulations and to motivate the prescription of stochastic processes in subgrid physics.

The most significant challenge in AGN studies is that each part of the electromagnetic spectrum captures a different aspect of the central engine, leading to major inconsistencies between detection methods \citep{2017A&ARv..25....2P}. This is particularly true for the obscured AGN population, nowadays suspected to also be a phase during the evolution of an AGN, and not only an outcome of a geometric alignment of the disc/torus system towards the observer.

The unification scenario \citep{1993ARA&A..31..473A, 1995PASP..107..803U} postulated that the observed variety in the presence of broad and narrow emission lines in the spectra of active galaxies was due to the obscuration induced by a molecular torus along the line of the sight of the observer. The distinction between AGN and quasars is largely a description of the relative luminosity of the central engine and the host galaxy, with quasars being extremely luminous and dominating over the host galaxy emission. The definition of red quasars corresponds to sources that show broad lines (i.e. type 1 sources) but with significant absorption in their continuum \citep{2012ApJ...757...51G, 2013MNRAS.429L..55B}. 
Red quasars seem to contradict the AGN unification scenario, as first \citet{2019MNRAS.488.3109K} and more recently { \citet{Andonie_2022}, \citet{2023MNRAS.525.5575F, 2024MNRAS.529.1995P, 2024A&A...691A.191C} and \citet{2024MNRAS.529.3939Y}, showed evidence of enhanced radio detection rates from this population, at odds with a simple orientation-induced obscuration.} Current models and several observations argue that red quasars could be an initial, short-lived stage during the onset of quasar activity within a galaxy. As the gas and dust is driven into the centre of the galaxy, the initial phase of accretion is enshrouded in a dusty cocoon. Subsequently, the radiation pressure, and induced winds will clear out the region around the black hole revealing a type 1, unobscured blue quasar \citep{2009ApJ...698.1095U, 2012MNRAS.427.2275B, 2015MNRAS.447.3368B, 2019MNRAS.487.2594T, 2021A&A...649A.102C}.

Determining the physical parameters of quasars and AGN and contrasting them with inactive galaxies (i.e., non-AGN hosts) as a function of luminosity, stellar mass, star formation rate, obscuration, as well as across cosmic time and large-scale environment is crucial \citep[e.g.,][]{2018MNRAS.475.3682W}, Laloux et al., in prep. It will enable the creation of an evolutionary scenario for galaxies including the incidence of AGN, which is much needed to establish the true evolutionary path of quasars and to inform the recipes used in simulations. However, AGN are short-lived phenomena and detailed statistical studies of this population have been hindered by the size of the available datasets. 

The value of large AGN and quasar samples is not only in the accurate determination of AGN luminosity distribution functions but also in allowing for detailed decomposition into AGN sub-populations. A recent decomposition of about $\sim 150$ mid-infrared detected AGN, split into unobscured, red quasars, and type 2 sources, shows the luminosity distribution functions to be a double-power law form with a break at a characteristic luminosity \citep{2018ApJ...861...37G}. Similar distributions are found also in other wavelengths \citep[e.g.,][]{2016A&A...587A.142F}. Much larger samples are needed to draw definitive conclusions, however these studies show already differences in the number density of red and blue quasars, or type 1 and type 2 quasars between low and high redshift ($z \sim 1$).

\Euclid is a mission of the European Space Agency (ESA), expected to detect billions of sources in the optical and near-infrared \citep{Laureijs11, EuclidSkyOverview}. The \Euclid observational campaign will observe a third of the extra-Galactic sky through two surveys. The Euclid Wide Survey (EWS), covering more than $\sim$ 14\,000 deg$^2$, and the Euclid Deep Survey (EDS), focusing on three different areas for a total of 63.1 deg$^2$. With its unprecedented, large dataset, \Euclid offers the possibility to study and identify the largest sample to date of unobscured and obscured AGN across all wavebands, extending to the faintest sources. 
A key aspect will be the determination of a corresponding selection function, which remains under ongoing development, particularly in the context of AGN studies.
This paper presents a study on red quasars within the \citet{Q1cite}, comparing mock and observed sources, and discussing and refining colour selection criteria that will be functional for the EWS.

\section{Dataset}
\label{sec:dataset}
\begin{figure*}
   \centering
   \includegraphics[width=\textwidth]{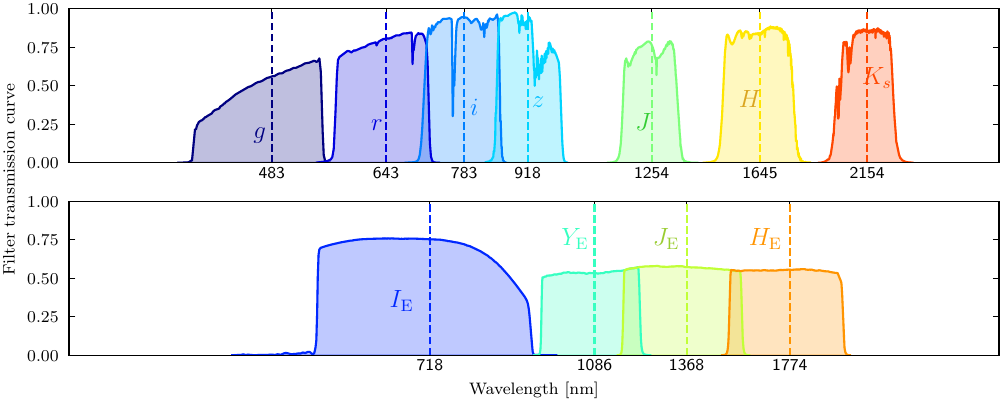}
   \caption{Set of passbands describing the properties of mock red quasars. The panels show the transmission efficiencies for the broad-band optical DECam $griz$ filters, the \Euclid VIS (optical) and NISP (NIR) filters, and the NIR to MIR VISTA $JHKs$ filters.
   The vertical dashed lines are placed at the weighted central wavelength of each filter.}
    \label{fig:01}
\end{figure*}

In the forecast analysis, we study the separation between mock red QSOs and observed sources. We use a synthetic catalogue for the former (more information follows in \cref{sec:mocks}) and the COSMOS2020 dataset \citep{2022ApJS..258...11W} for the latter. Mock red quasars are described over a range of wavelengths extending from the optical through the near infrared (NIR), and up to the mid-infrared (MIR): DECam $g, r, i, z$ in the optical \citep{10.1093/mnras/stw641}, \Euclid VIS (\IE) and NISP \YE, \JE, \HE in NIR \citep{EuclidSkyVIS, EuclidSkyNISP, Schirmer-EP18}, VISTA $J, H, K_s$ from NIR to MIR \citep{2012A&A...544A.156M} and WISE $W1, W2$ in MIR \citep{2010AJ....140.1868W}. The collection of filters used in this work and their weighted central wavelengths is shown in \cref{fig:01}. The employed photometry is expressed in the AB magnitude system.

We clean the COSMOS2020 sample from corrupted photometry and fake detections, and exclude the objects with null entries for the photometric redshift. We apply the mask $\texttt{FLAG\_COMBINED}=0$ to remove objects near bright stars and saturated regions, $\texttt{ACS\_MU\_CLASS}=1$ to separate stars from galaxies, and $\texttt{lp\_type}\neq 9$ to exclude fake detections. Furthermore, we apply a magnitude upper cut at 23.5 in VISTA $H$ in both COSMOS2020 and the mocks. Among the selected sources in the COSMOS2020 catalogue, we take into account those that are flagged as AGN candidates. The selected catalogue includes 95\,052 objects. 

Additionally, we keep track of 1493 X-ray sources observed with \Chandra \citep{2016ApJ...819...62C}, which we do not consider in this analysis. X-ray-selected AGN outline a broad range of obscuration, and their optical-to-NIR colours may not fully align with the selection criteria used in this study, based solely on red QSO NIR photometry. We will address this additional level of complexity in future work, using their properties to refine the distinction between reddened AGN and red galaxies.

We analyse mocks and COSMOS2020 (methods are described in \cref{sec:methods}) to derive a selection function for candidate red QSOs, and we apply it to \Euclid Q1 \citep{EuclidSkyOverview}. Q1 consists of a first visit of the Euclid Deep Fields (EDFs), spanning across a total area of $63.1\,\mathrm{deg}^2$ of the extragalactic sky, divided in the Euclid Deep Field North (EDF-N, $20\,\mathrm{deg}^2$), Euclid Deep Field Fornax (EDF-F, $10\,\mathrm{deg}^2$) and the Euclid Deep Field South (EDF-S, $23\,\mathrm{deg}^2$). More details about the Q1 release are presented in \citet{Q1-TP001, Q1-TP002, Q1-TP003} and \citet{Q1-TP004}. 

In this work, we focus on the EDF-F, for which we find overlap with a collection of AGN candidates from \citet{2022ApJS..262...15Z}, selected through the flag $\texttt{flag\_IRagn\_D12}$, which follows the MIR colour-based cut proposed in \citet{2012ApJ...748..142D}, and a catalogue of radio-selected quasars \citep{2013ApJS..205...13M}. These two datasets serve as control samples. We build a first dataset of 5\,301\,332 EDF-F sources, obtained by matching the \Euclid morphology \citep{Q1-TP004} and photometric redshift \citep{Q1-TP005, Q1-SP026} catalogues, delivered by the OU-MER and OU-PHZ organizational units, respectively. Then we exclude objects flagged as spurious and with unphysical photometric and redshift properties. Furthermore, we consider only sources classified either as galaxies or QSOs, according to the OU-PHZ classification presented in \citet{Q1-SP026}. A summary of this selection function is reported in \cref{tab:03}. Additionally, we apply a cut near the limiting magnitude of the \HE band, corresponding to 23.5. The final subsample at play counts 1\,331\,325 sources.

\begin{table}
\caption{Sample selection function applied to the Q1 sample. In $\mathrm{FLUX\_filter\_nFWHM\_APER}$, $\mathrm{filter}$ is the passband and 
$n=1,2,3,4$
according to the aperture.}
\centering
\begin{tabularx}{\columnwidth}{Xcc}
\toprule
\textbf{Feature in \Euclid Q1} & \textbf{Selected values} \\
\midrule
$\mathrm{PHZ\_CLASSIFICATION}$       & [2,6] \\
$\mathrm{PHZ\_MEDIAN}$               & finite \\
$\mathrm{SPURIOUS\_FLAG}$            & 0      \\
$\mathrm{DET\_QUALITY\_FLAG}$        & $<8$   \\
$\mathrm{FLUX\_filter\_2FWHM\_APER}$ & $>0$   \\
\bottomrule
\end{tabularx}
\label{tab:03}
\end{table}

\subsection{Template fitting}
\label{subsec:TF}
Each object in COSMOS2020 is described by a photometric dataset including ultraviolet (UV) measures from GALEX, optical observations from the Subaru Hyper Suprime-Cam (HSC) and the Canada-France-Hawaii Telescope (CFHT), NIR data from VISTA, MIR  from the SPLASH program of the \Spitzer Space Telescope, and optical medium band observations from Subaru. We refer to \cite{2022ApJS..258...11W} for references on these individual datasets. The properties of each filter are described in \cref{tab:filters}. 

The mock dataset includes VISTA, DECam, and \Euclid VIS and NISP photometry. In order to match the photometric datasets of mocks and COSMOS2020, we use template fitting (TF) to estimate the optical and NIR fluxes that are originally not available in the latter. 
TF compares the input photometric dataset with a library of Spectral Energy Distributions (SEDs) to identify the best match. For this task we use the \texttt{Phosphoros} package (Paltani et al., in prep). \texttt{Phosphoros} is a fully Bayesian TF algorithm, supporting flexible prior distributions across all parameters (redshift, reddening, 
SED-index and luminosity) and producing multi-dimensional and marginalized posterior distributions. It was successfully employed in the \Euclid photo-z challenge, presented in \citet{2020A&A...644A..31E}, which was designed to evaluate the accuracy of various methods for photometric redshift estimation against the stringent requirements of \Euclid cosmic shear analyses (\citealt{2013MNRAS.431.3103C}, Tarsitano et al., in prep.). In \cite{2023A&A...670A..82D}, \texttt{Phosphoros} was validated and benchmarked against a similar code, \texttt{Le Phare} \citep{2011ascl.soft08009A}. For each input galaxy, \texttt{Phosphoros} provides a multivariate posterior distribution, allowing the inference of flux estimates in the missing bands from the best fit. 
For additional details about its metrics and models we refer the reader to \cite{2023A&A...670A..82D} and \citet{Q1-SP026}.

\subsection{Mock red QSO catalogue}
\label{sec:mocks}

A sample of SDSS QSOs was selected across redshift and grouped into nine bins of FWHM and equivalent width. Stitching together the stacked spectra, we created a composite spectrum with very broad wavelength coverage. 

\begin{figure*}
    \centering
    \includegraphics[width=\textwidth]{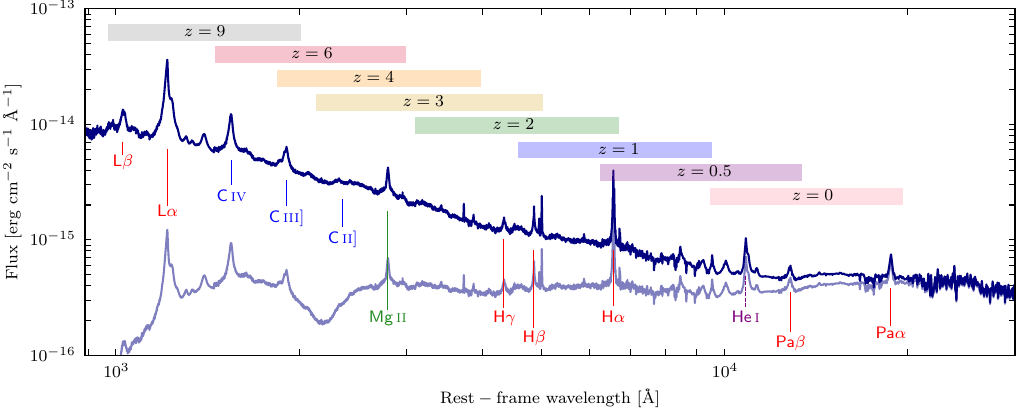}
    \caption{Spectrum comparison between the unobscured QSO stack \citep[dark blue line;][]{EP-Lusso}, and a reddened QSO spectrum with $E(B-V)=0.25$ (bright blue line). Emission lines are highlighted using coloured vertical markers along with their respective labels, pointing to different atomic species and ionization states. The coloured bars correspond to the redshift range of the red-grism of NISP.}
    \label{fig:wp9-qso}
\end{figure*}

\Cref{fig:wp9-qso} shows the unobscured QSO spectrum and the dramatic impact of $E(B-V)=0.25$ attenuation applied to it. The coloured bars correspond to the rest-frame wavelength coverage of the red-grism of \Euclid. To create a mock catalogue of red QSOs, we used the first bin of the stacked QSO spectra of \citet{EP-Lusso} and the luminosity function of red QSO determined in \citet{2018ApJ...861...37G}. The latter is described by a double power-law function, already presented in \citet{2015ApJ...802..102L}, characterised by a faint-end and a right-end slope, and a break luminosity, where the dominance shifts from the faint to the bright end.

We created a grid of bolometric luminosity ($40<\logten L_{\rm bol}<50$) and redshift ($0<z<8$), and calculated the expected number of red QSOs by integrating the luminosity function. For each mock SED, we applied reddening according to their distribution covering $0.25<E(B-V)<1.45$, and assuming the Prevot attenuation law and intergalactic medium attenuation as described in \citet{EP-Lusso}. 

\begin{figure}
    \centering
    \includegraphics[width=\linewidth]{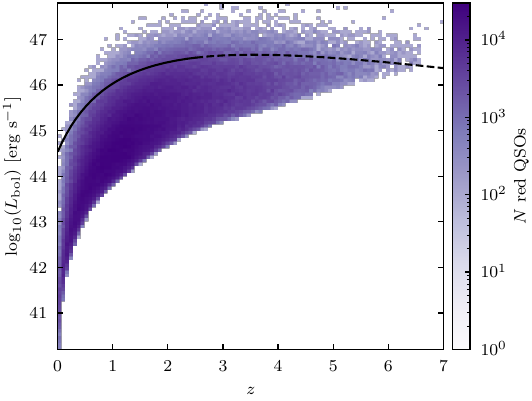}
    \caption{Luminosity-redshift plane for the mock sample of red QSO. The black line is the knee of the luminosity function of \citet{2018ApJ...861...37G}.}
    \label{fig:mock-J-LF}
\end{figure}

Finally, we applied an observed magnitude cut corresponding to the expected depth of EWS, i.e. $J<24.5$. \Cref{fig:mock-J-LF} shows the coverage of the luminosity-redshift plane of our mock catalogue, assuming 14\,500\,$\rm deg^2$ sky coverage, at the wide-depth of \Euclid. The black line shows the break luminosity of red QSOs from  \citet{2018ApJ...861...37G}.


\section{Methods}
\label{sec:methods}
 Colour-based selection criteria have been extensively studied in literature to identify AGN and reddened AGN. Most notably, \citet{2004ApJS..154..166L}, \citet{2005ApJ...631..163S}, \citet{2012hcxa.confE.148M} and \citet{2018ApJS..234...23A} defined empirical cuts in MIR colour space. Alongside MIR-based selections, NIR colours have been proposed in \citet{2012MNRAS.427.2275B} and \citet{ 2012ApJ...757...51G, 2013ApJ...778..127G, 2018ApJ...861...37G}, to separate red quasars from stars and galaxies. 

To follow-up on these studies exploiting the unique depth and high-resolution of \Euclid, we introduce a novel selection method that is solely based on \Euclid NIR photometry. Our goal is to enhance the systematic identification of red QSOs in the EWS where MIR photometry may be incomplete or unavailable. 

To achieve this, we conduct a series of statistical analyses to study the photometric selection criteria that maximise the distance between the COSMOS2020 observed dataset and the mock red QSOs. In particular, we explore the separation in a multi-dimensional colour-colour space, as a function of magnitude and redshift. Our study is performed independently using VISTA+DECam and reconstructed \Euclid photometry. This allows us to assess the impact of different wavelength coverage and filter sets on the identification of red QSOs, and to evaluate the \Euclid NIR stand-alone capability in recovering our target population.

First, we collect features that are directly transferable into the reconstructed \Euclid-like photometry. More precisely, we consider VISTA $J - K_s$, $Y - K_s$ and $J - H$ to describe NIR colours, and \mbox{DECam} $i - K_s$ to sharpen the selection of reddened sources based on their optical-to-NIR transition. In the \Euclid-like colour space, we use $\YE - \HE$, $\JE - \HE$, and $\YE - \JE$ as NIR colours, and we adopt $\IE - \HE$ as a metric to estimate the optical-to-NIR excess. We proceed with the implementation of a multi-step analysis:

\begin{enumerate}[i]
    \item Principal Component Analysis (PCA): this method has been successfully applied in previous studies to investigate the underlying physical properties of AGN and their host galaxies. \citet{1992ApJS...80..109B} were the first to use it in the AGN domain, to analyse optical emission-lines and continuum properties of a low-redshift quasar sample. Their work was followed by \citet{1993ApJ...403L...9C}, \citet{1994ApJ...435..611L, 1997ApJ...477...93L}, \citet{1996A&A...309...81W}, and \citet{1998AN....319....7B}. Their analyses found correlation between the primary eigenvector (or principal component) and quasar spectral features depending on physical parameters including Eddington ratio, luminosity and black hole spin. Beyond optical emission lines, PCA has been applied to AGN spectral energy distributions and photometric datasets. \citet{2004AJ....128.2603Y} applied PCA on SDSS quasar spectra, and showed that the eigenvectors (named \textit{eigenspectra} in their work) have the power of disentangling the contribution of the host galaxy light, the optical continuum and the AGN emission. \citet{Kuraszkiewicz_2009} analysed a sample of red 2MASS AGN \citep{2002ASPC..284..127C}. Among their results, they found that the second principal component was correlated with optical-to-infrared colours ($B - K_s, B - R, J - K_s$), depending on the contribution of the host galaxy relative to the AGN emission. PCA was used to study AGN samples also in \citet{Hao_2005}, \citet{2012MNRAS.423..600S}, \citet{2024ApJS..272...13P}.
    
    In this work, we apply PCA to the aforementioned multidimensional colour space to identify the most informative colours that separate mock red QSOs from the observed COSMOS2020 sources. Our analysis focuses on broadband photometric selection in preparation for systematic large-scale red QSO searches in the \Euclid Survey. Through linear combination of the original features, PCA reduces dimensionality while capturing variance in the dataset. It serves as an exploratory framework to highlight which optical and NIR colours contribute the most to the identification of red QSOs. Detailed information follows in \cref{subsec:methods_PCA}.
    \item Empirical colour-colour cuts: using the most significant colours identified via PCA, we study a colour-colour selection function for red QSOs. 
    Additional information on the metric adopted to evaluate the selection performance is reported in \cref{subsec:methods_SA}.
    \item Machine learning-based refinement: in this phase, we train a probabilistic Random Forest classifier (RF, \citealt{10.1023/A:1010933404324}) to refine the previous selection function for red QSOs and mitigate the effects of contaminants. RF has the advantage of handling non-linear relationships between the input features, so it sets complex decision boundaries that PCA and empirical colour-based cuts cannot capture. The trained RF model is then applied to the \Euclid Q1 dataset, where we select candidate red QSOs based on their predicted probabilities. We refer the readers to \cref{subsec:method_RFC} for a detailed description of this method.
\end{enumerate}

In the analysis, we consider the discriminating power of additional features, such as compactness criteria, and we use external AGN datasets as control samples. The latter play a crucial role in identifying the proposed selection function and assessing its robustness, or revealing the risk of introducing a bias that could reduce the completeness of the red quasar selection. We use our findings to build a first census of candidate red QSOs in \Euclid Q1 EDF-F. Such a sample provides a framework for future spectral analyses and will serve as a training set for Artificial Intelligence-based automated classification, extended to the EDF-N, EDF-S, and the 
EWS.

\subsection{Hyper-colour determination and colour selection}
\label{subsec:methods_PCA}

Principal Component Analysis (PCA) is a statistical method that can be used to project a dataset from a high-dimensional space into a low-dimensional space, retaining its most meaningful properties. More precisely, the technique consists of mapping the original features into a new set of uncorrelated ones, named principal components. The result is a linear combination where each coefficient represents the contribution of the corresponding original feature in forming a principal component. With this technique, the dataset is linearly transformed onto a new coordinate system whose directions, the principal components, capture as much variance in the data as possible, with the first component capturing the most variance, the second capturing the next most, and so on.

In this work, we use the module \texttt{PCA} available in the \textit{Python} library \texttt{Scikit-learn} \citep{scikit-learn}, which follows the implementation presented in \cite{2009arXiv0909.4061H}. Applying PCA decomposition to our multi-dimensional dataset, described by the aforementioned colour features, means collapsing it into a low-dimensional one, where each principal component (PC) is a linear combination of the original colours ($c$) as:

\begin{equation}
    \mathrm{PC} = \sum_{j=1}^{N} a_{j} c_{j} \ ,
    \label{eq:03}
\end{equation}

\noindent
with $a_{j}$ the coefficients of the linear combination and $N$ the number of involved colours. Features with consistently high coefficients across components explain a larger portion of the variance and are more important in the transformed space. The aim is to identify principal components receiving significant contributions by a set of original colours, and study their discriminating power between red QSOs and the rest of the sample. We will refer to such principal components as \textit{hyper-colours} (HC). 

We run PCA on the mocks and COSMOS2020, standardising the input features to ensure comparability across different scales. By assuming three principal components, and we identify a cut in the HC space which guarantees the highest discriminating power. We will refer to this cut as hyper-colours cut (HP-cut). 

Furthermore, we study the impact that the single colour features, $c_{j}$, have on the HC, based on their linear coefficients $a_{j}$, and we identify a cut in the multi-dimensional colour-colours space made by them. We will refer to this selection cut as colour-colour cut (cc-cut).

\subsection{Forecast analysis}
\label{subsec:methods_SA} 

By applying the HP- or cc-cut, we estimate the completeness ($C$) and purity ($P$) of the selected sample.
Completeness is the fraction of mock red quasars correctly identified by the proposed selection criterion and is defined as:

\begin{equation}
    C \mathrm{\ = \frac{TP}{TP + FN}} \ ,
    \label{eq:01}
\end{equation}

\noindent
where TP (True Positives) and FN (False Negatives) are the number of red quasars correctly identified and missed by the cut. The sum $\mathrm{TP+FN}$ then corresponds to the total of red quasars in the mock sample.
Purity is defined as the fraction of TP among all selected objects. We calculate it as:

\begin{equation}
    P\mathrm{\ = \frac{TP}{TP + FP}} \ ,
    \label{eq:02}
\end{equation}

\noindent
where FP (False Positives) is the number of sources incorrectly identified as red quasars by the selection cut. In this work, we define FP as the number of COSMOS2020 selected sources which are not classified as AGN, plus the number of AGN candidates passing the cut with DECam $i - K_s < 1.7$. We assume this metric since $1.7$ is the lower limit for optical-to-NIR excess in the mock sample. Finally, we apply to EDF-F the colour-based selection functions, and we study them with the aid of two control samples of MIR-selected and radio-selected AGN candidates in the same field. 

\subsection{Probabilistic Random Forest}
\label{subsec:method_RFC}
A Random Forest is a machine learning algorithm that consists of creating an ensemble of decision trees and combines their outputs to make predictions. Each tree in the forest is trained on a random subset of the data and features, and the final classification is made by majority voting (standard RF), or by averaging the predicted probabilities assigned to each class across all the decision trees (probabilistic RF). In a binary classification case, probabilistic RF assigns each source two values, corresponding to the probability of belonging to each of the two classes. This approach allows us to estimate the confidence level of each classification and make a probabilistic selection of candidate red QSOs.
RF can identify and rank the most important features that differentiate red quasars from other objects, and it can handle complex and non-linear relationships between features. Furthermore, RF is more robust towards over-fitting, which makes it suitable for noisy or imbalanced datasets like in the case of targeting red QSOs \citep{10.1023/A:1010933404324, chen04using}. In our work, we used the module \texttt{RandomForestClassifier} available in the \textit{Python} library \texttt{Scikit-learn}. 

First, we train a RF classifier on three sets of features (named \texttt{S1}, \texttt{S2} and \texttt{S3}) based on different combinations of \Euclid-based colours and magnitudes. More precisely, \texttt{S1} includes the most significant colours identified by PCA and \texttt{S2} all the \Euclid NIR colours. \texttt{S3} adds \Euclid magnitudes on \texttt{S2}. This multi-steps approach allows us to test the impact of expanding the primary feature set in terms of purity, completeness and classification performance. 

For each set of features we identify the best model using hyper-parameter tuning.
The RF model operates within an hyper-parameter space described by a set of key parameters, whose combinations can impact the performance of the model itself. These parameters include the number of trees, the minimum amount of samples required to split a tree node and the leaf size. We used the \texttt{Scikit-learn} module \texttt{RandomizedSearchCV} to explore this hyper-parameter space and identify the best performing set of key parameters. For each sampled combination of parameters, the algorithm employs a stratified k-fold cross-validation strategy with $k=5$ folds. This technique divides the dataset into five partitions and trains the model on four of them, using the fifth for validation. The goal of this approach is to get a robust estimate of the model performance and check if over-fitting occurs to certain partitions. The best combination of parameters is selected based on the cross-validation accuracy. 

Through hyper-parameter tuning and cross-validation (run with a fixed random seed for reproducibility), we obtain an optimised RF model for each initial set of features (\texttt{S1}, \texttt{S2} and \texttt{S3}). We compare the three models in terms of feature importance, completeness and purity, and we apply the best one to the \Euclid Q1 EDF-F dataset. We discuss the results and our findings in \cref{sec:results}.

\section{Results}
\label{sec:results}

In this section, we present the findings we obtained for the steps described in \cref{sec:methods}.

\subsection{Selection function in the hyper-colour space}
\begin{figure}
   \centering
   \includegraphics[]{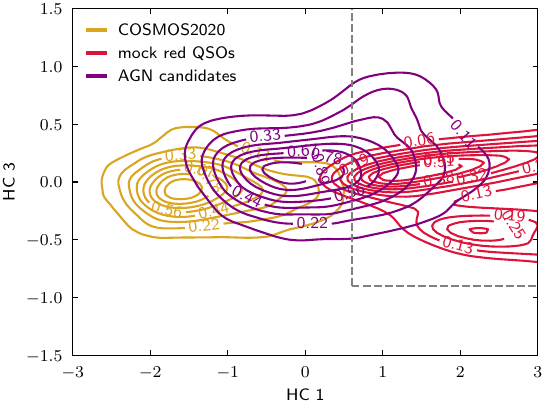}
   \caption{Hyper colour-colour diagram showing the separation of mock red QSOs from COSMOS 2020 sources. The contours are normalised to their respective subsets. The proposed selection function is displayed by the gray dashed lines.}
    \label{fig:img_03}
\end{figure}

The PCA on mocks and COSMOS2020, run in VISTA-defined multidimensional colour space, provides three HC whose coefficients are reported in \cref{tab:pca_coefficients_uvista}. HC1 is a weighted average of all the input colour features, with a slight emphasis on $J-K_s$ and $i-K_s$. HC2 and HC3 are dominated by $H-K_s$ and $i-K_s$. Their explained variances (91\%, 6.2\% and 2.7\%, respectively) indicate that HC1 is related to the overall colour gradients across the feature set, while HC2 and HC3 isolate information specific to a certain feature.
The combination of HC1 and HC3 provides the strongest discriminating power, as displayed in \cref{fig:img_03}. The contours, normalised to their respective subsets, show that this HC space is able to effectively disentangle the populations of observed sources and mock red QSOs. The HC-cut of $\mathrm{HC1 > 0.6}$ and $\mathrm{HC3 > -0.9}$ leads to an overall completeness of 98\% with 81\% purity. Among the AGN candidates passing the cut, 91\% of them have DECam $i - K_s > 1.7$. As reported in \cref{sec:methods}, based on mocks we consider this as a requirement to identify possible obscured AGNs and flag them as candidate red QSOs. 

\begin{figure}
   \centering
   \includegraphics[]{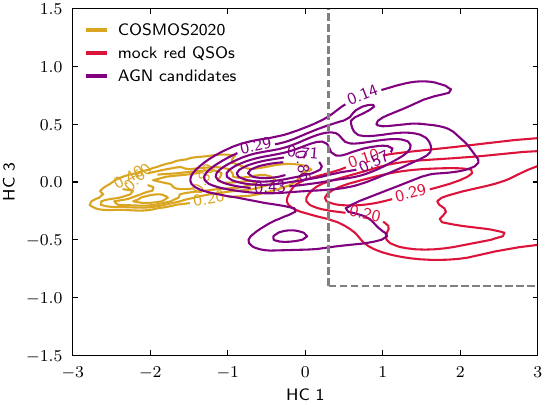}
   \caption{Hyper colour-colour diagram obtained using \Euclid-like photometry, showing the separation of mock red QSOs from COSMOS 2020 sources. The proposed selection function is displayed by the gray dashed lines.}
    \label{fig:img_04}
\end{figure}

We run again the PCA, transferring the original VISTA colour features into the \Euclid-like colour space, yielding to similar results. The coefficients of the principal components are displayed in \cref{tab:pca_coefficients_euclid}.
Analogously to the VISTA-based HC, the \Euclid-based HC1 is dominated by $\YE - \HE$ and HC3 gets most contribution from $\IE - \HE$, tracking the optical-to-NIR transition. \Cref{fig:img_04} displays the \Euclid-like HC space formed by HC1 and HC3. An HC-cut of $\mathrm{HC1 > 0.3}$ and $\mathrm{HC3 > -0.9}$ leads to an overall completeness of 97\% with 68\% purity.  

Beyond unveiling hyper-colours, PCA highlights the original colour features, $c_{j}$ in \cref{eq:03},  bringing the most weight in the identification of candidate red QSOs. More precisely, in the VISTA parameter space, the most important colours for HC1, HC2 and HC3, based on their linear coefficients, are $J-K_s$, $H-K_s$ and $i-K_s$. In the \Euclid-like parameter space, the most important features for the tree components are $\YE - \HE$, $\JE - \HE$ and $\IE - \HE$, respectively. We further advance our analysis by a visual and quantitative assessment of the separation of mock red QSOs using these multidimensional colour-colour spaces.

\begin{figure}
   \centering
   \includegraphics[]{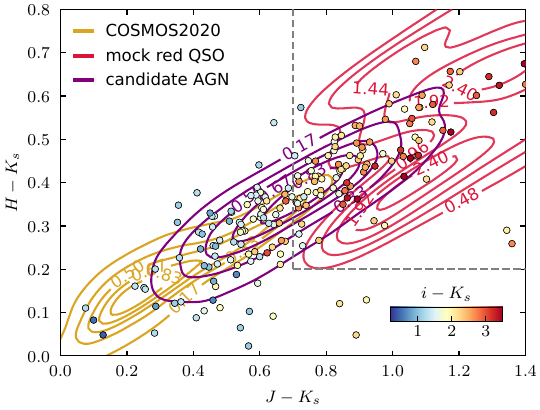}
   \caption{The VISTA colour-colour diagram showing the separation of mock red QSOs from COSMOS 2020. The proposed colour-colour selection function is displayed by the gray dashed lines. A subset of candidate AGN overlays the selection, colour-coded as $i - K_s$.}
    \label{fig:img_cc_03}
\end{figure}

\begin{figure}
   \centering
   \includegraphics[width=\columnwidth]{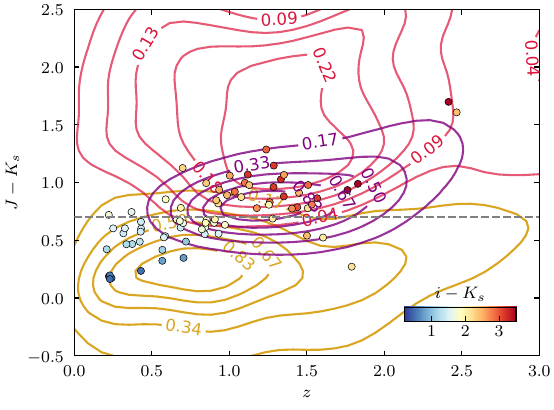}
   \caption{The VISTA colour-$z$ diagram showing the separation of mock red QSOs from COSMOS2020. The colour-code adopted for displayed populations are as in \cref{fig:img_cc_03}.}
    \label{fig:img_06}
\end{figure}

\begin{figure}
   \centering
   \includegraphics[]{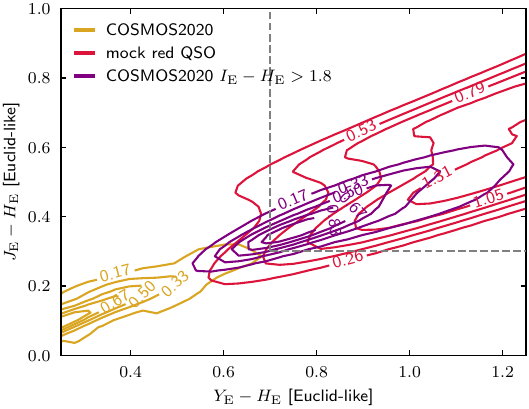}
   \caption{The \Euclid-like colour-colour diagram showing the separation of mock red QSOs from COSMOS2020. The proposed colour-colour selection function is displayed by the gray dashed lines. The COSMOS2020 passing the selection is displayed in purple.}
    \label{fig:img_05}
\end{figure}

\begin{figure}
   \centering
   \includegraphics[]{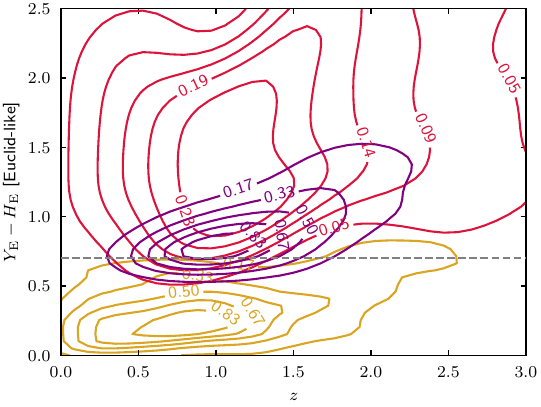}
   \caption{The \Euclid-like colour-redshift diagram showing the separation of mock red QSOs (in red) from COSMOS2020 (golden contours). The COSMOS2020 passing the proposed cc-cut is displayed in purple.}
    \label{fig:img_08}
\end{figure}

\subsection{Photometric selection in colour-colour space}
\label{subsec:results_colours}
\Cref{fig:img_cc_03} displays the separation between mock red QSOs and the COSMOS2020 dataset achieved in a multidimensional parameter space spanning from optical to NIR wavelengths and defined through the most important colour features according to PCA. The two populations are compared in the $J-K_s$ vs $H-K_s$ diagram, and COSMOS2020 is further split in two subsets to highlight candidate AGN.
If we apply the cut $J-K_s > 0.8$ and $H-K_s > 0.2$, we obtain an overall completeness of 99\% and a purity of 78\%. The 85\% of the candidate AGN passing this multidimensional colour cut responds to the mock-calibrated red excess of DECam $i - K_s > 1.7$. 
As the cc-cut is defined through optical and NIR colours, the estimate of purity can be interested by dependences on redshift and magnitude. 
We first estimate it as a function of redshift, identifying intervals with major degeneracies between the populations at play (\cref{fig:img_06}). We obtain $90 \% $ purity at $\mathrm{\textit{z} <0.5}$, $69 \% $ at $\mathrm{0.5 < \textit{z} <1.5}$, and $87 \% $ at higher redshifts. The magnitude dependency sees purity values of 77\% for $H<20$, 65\% for $20< H <22$ and 86\% for objects in the interval $22 < H < 23.5$.

We proceed with the forecast analysis using the same metric, but defined with \Euclid-like photometry. We study a grid of multidimensional cc-cuts to maximise the separation between the subsets at play, and we weight the FP rate with the unreddened COSMOS2020 AGN candidates. The results are displayed in \cref{fig:img_05}, with the colour-redshift evolution shown in \cref{fig:img_08}. In this case we find that the colour cut, $\YE - \HE > 0.7$ and $\JE - \HE > 0.3$ with $\IE - \HE > 1.8$, leads to an overall 99\% completeness and 67\% purity, with redshift-dependent fluctuations: 86\% purity at $z <0.5$, 57\% and 63\% at $0.5 < z <1.5$ and higher redshifts, respectively. Purity values with magnitudes are 56\% for $\HE <20$, 50\% for $20 < \HE < 22$  and 77\% for objects in the interval $22 < H < 23.5$.

\subsection{Colour-colour selection applied to \Euclid Q1}
\label{subsec:ResEuclid}
\begin{figure}
   \centering
   \includegraphics[]{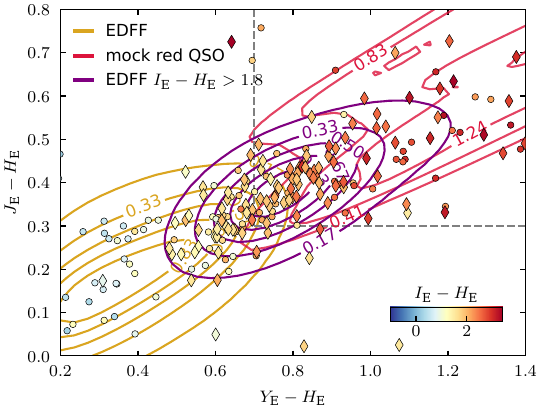}
  \caption{The colour-colour diagram showing the separation of mock red QSOs from the \Euclid Q1 sample. The proposed colour-colour selection is displayed by the grey dashed lines. The control samples are colour-coded by $\IE - \HE$: dots and diamonds represent MIR-selected and radio-selected AGN, respectively.}
    \label{fig:img_09}
\end{figure}

\begin{figure}
   \centering
   \includegraphics[]{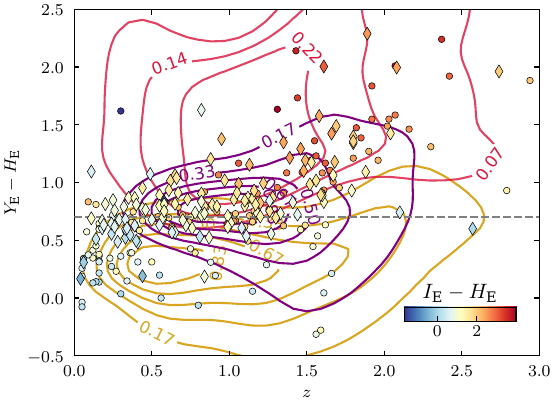}
   \caption{The colour-$z$ diagram showing the separation of mock red QSOs from the \Euclid Q1 sample. The proposed colour-colour selection is displayed by the grey dashed line. The control samples are colour-coded by $\IE - \HE$: dots and diamonds represent the MIR-selected and the radio-selected AGN, respectively. The contours are coloured as in \cref{fig:img_09}.}
    \label{fig:img_10}
\end{figure}

Having established the methods for the selection of red quasars based on mock and observed training samples in the previous sections, we now apply these methods to the \Euclid Q1 EDF-F dataset.
The selection is supported by control samples, MIR-selected and radio-selected AGN candidates, introduced in \cref{sec:dataset}. \Cref{fig:img_09} displays the discriminating power of the proposed multidimensional colour-colour cut, yielding to an overall 98\% completeness of the mock sample. The colour-redshift evolution is displayed in \cref{fig:img_10}. The MIR-selected and radio-selected AGN have a percentage of 77\% and 47\% reddened sources, according of the proposed cut $\IE - \HE > 1.8$. Among them, the 87\% passes the multidimensional cc-cut.
The selection functions are summarized in \cref{tab:02}. Combining the cc-cut with the RF classification probability (described in \cref{sec:methods}) we flag 151\,853 sources as candidate red QSOs.

\begin{table}
\centering
\caption{Forecast red QSOs Completeness ($C$) and Purity ($P$) from NIR selection functions. The first two rows report the selection function defined in the VISTA colour space, the last two refer to the reconstructed \Euclid photometry.}
\begin{tabularx}{\columnwidth}{Xcc}
\toprule
\textbf{Candidate red QSO selection function} & \textbf{$C$} & \textbf{$P$} \\
\midrule
$\mathrm{HC1 > 0.6 \ and \ HC3 > -0.9}$            & 0.99                 & 0.82           \\[0.1cm]
$J - K_s > 0.7, \ H - K_s > 0.3$, & \multirow{2}{*}{0.99} &  \multirow{2}{*}{0.79} \\
and $i-K_s > 1.7$ \\[0.1cm]
$\mathrm{HC1 > 0.3 \ and \ HC3 > -0.9}$ [\Euclid-like]            & 0.97                 & 0.78           \\[0.1cm]
$\YE - \HE > 0.7, \ \JE - \HE > 0.3$, &  \multirow{2}{*}{0.98} &  \multirow{2}{*}{0.80} \\ 
and $\IE - \HE > 1.8$ \\
\bottomrule
\end{tabularx}
\label{tab:02}
\end{table}

 \subsection{Random Forest analysis}
\label{subsec:ResEuclid_RF}
We apply a probabilistic RF classifier to identify candidate red quasars starting from a first set of photometric features, \texttt{S1}, including the most significant colours according to PCA: $\JE - \HE$, $\YE - \HE$ and $\IE - \HE$. The hyper-parameter search lead to an optimised model with 100 trees, minimum split size of 10 and minimum leaf size equal to 1. This configuration yields to a mean cross-validation accuracy of $95 \%$ with a standard deviation of $< 1 \%$. We recover feature importance values of 0.35 for $\JE - \HE$, 0.39 for $\YE - \HE$ and 0.26 for $\IE - \HE$. Applying a probability threshold of $P > 0.7$ for classification, we obtain $98 \%$ completeness and $87 \%$ purity. We determine the probability threshold so that the completeness does not fall below the value we achieved using empirical cc-cuts. This criterion ensures that the RF classification is conservative at least as much as the other selection methods while improving purity. 

\begin{figure}
   \centering
   \includegraphics[]{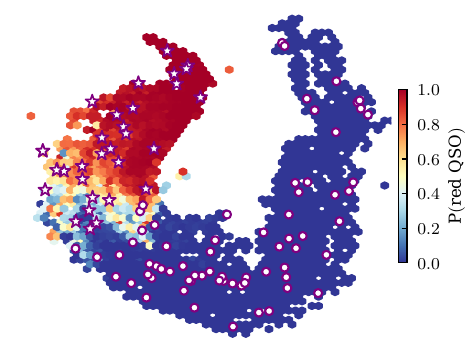}
   \caption{UMAP visualisation for the classification results of the probabilistic RF run on the most significant colour features. Hexagons are colour-coded by the probability of being a candidate red quasar. The overlayed scatter plot displays a random subset with symbols reflecting the empirical cc-cut. Stars represent objects previously classified as red QSOs. Circles are employed otherwise.}
    \label{fig:UMAP_S1}
\end{figure}

To gather further insights from our analysis, we employ the Uniform Manifold Approximation and Projection algorithm (UMAP). As described in \citet{2018arXiv180203426M}, UMAP is a non-linear dimensionality reduction technique that preserves both the local and global structure of the data and highlights possible clusters and patterns. We use UMAP to visualise the dataset in a reduce-dimensionality space and check the separation between classes. The UMAP visualization of the classified objects (\cref{fig:UMAP_S1}), colour-coded by the probability of being a red QSO, shows that a threshold of 0.7 outlines a boundary region between the two classes. One-hundred sources randomly drawn from the test set are plot over the map, with symbols following the empirical cc-cut: stars for sources classified as red quasars, circles for sources that did not pass the selection. The overlay displays a direct comparison between the empirical colour selection method and the RF classification. Previously selected sources are mostly present in the region with higher probability of being a red quasar, and a minority populates the transition region. The RF refinement of these boundaries goes beyond the level of accuracy achieved by the empirical cc-cut, thereby reducing contamination and raising purity.

We repeat the analysis on extended sets of features, specifically on \texttt{S2} and \texttt{S3}, in order to assess the impact of additional information on the classification performance. More precisely, \texttt{S2} includes $\YE - \JE$, $\JE - \HE$, $\YE - \HE$ and $\IE - \HE$, while \texttt{S3} expands \texttt{S2} with the magnitudes $\IE$, $\YE$, $\JE$ and $\HE$. 
The RF classifier optimised for \texttt{S2} yields to similar results, without improving completeness and purity. The feature importance analysis assigns to the additional colour, $\YE - \JE$, a value of 0.03. 
Such results aligns with our findings from \texttt{S3}. In this case, the importance of the magnitude features is $< 2\%$ and no increase in completeness and purity is registered. 

Along this multi-step analysis, the dominant importance of the PCA-selected colours remains unchanged and leads to a consolidated probabilistic RF model that we apply to the \Euclid Q1 EDF-F sample.

\begin{figure}
   \centering
   \includegraphics[]{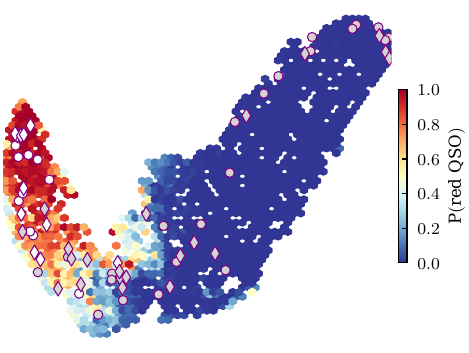}
   \caption{UMAP visualisation of the \Euclid Q1 dataset, colour-coded by the RF-based probability of being a red QSO. The overlayed scatter plot displays the $25 \%$ of the validation samples. MIR and radio-selected candidate AGN are marked by circles and diamonds, respectively. The symbols are filled in white if the source passed the empirical cc-cut, otherwise they are coloured in light gray.}
    \label{fig:UMAP_RF}
\end{figure}

The UMAP visualisation of the \Euclid Q1 sample, displayed in \cref{fig:UMAP_RF}, is colour-coded by the probability of being a red QSO. The over-plotted symbols represent a fraction ($25 \%$) of the two control samples: circles denote the MIR-selected AGN candidates, while diamonds represent the radio-selected AGN candidates. White markers indicate the sources that passed the previous empirical cc-cut selection, and light gray colours those that were not classified as red quasars. Analogously to our findings from the analysis of the COSMOS2020 dataset, we notice that spatial distribution of the sources clusters around two different populations, according to their likelihood of being red QSOs, and that the boundaries obtained with RF are more robust against contaminants, compared to the empirical cc-cuts.

\section{Discussion}
\label{sec:Discussion}
The VISTA-based HC spaces provide effective separation between mock red QSOs and observed sources. Their translation into the \Euclid-like HC space preserves the structure of the primary component (HC1), while introducing small shifts in the secondary components (HC2 and HC3). These shifts are expected due to the differences in filter characteristics and photometric uncertainties. The photometric selection function, using multi-dimensional colour cuts in the \Euclid optical and NIR regimes, forecasts high completeness (98\%) and moderate purity (78\%) for candidate red QSOs. Completeness of the mocks remains consistently high across redshift and magnitude intervals, suggesting that the proposed selection effectively captures the reddened QSO population. Purity decreases in intermediate redshift ranges and at faint magnitudes, where physical degeneracies between red QSOs and red galaxies become more significant. However, comparing redshift and magnitude distributions of the mock red QSOs and the sources in COSMOS2020 suggests there is also an observational effect at play. \Cref{fig:A1} shows that the redshift distribution of mock red QSOs peaks at higher values compared to the observed sources, as they are modelled to represent a deeper survey aiming to capture the properties of obscured AGN. Purity estimates can be affected in the mismatched higher-redshift range. The lower purity at intermediate redshift could be partially mitigated by training the selection function on larger, more representative datasets that include additional sources of variation, such as differences in host galaxy properties or environmental factors. Future work could explore the inclusion of environmental parameters, such as local density or clustering, to refine selection criteria. Larger values of purity for bright objects is also influenced by the broader photometric extent of \Euclid-like mocks, which outnumber COSMOS2020 at lower magnitudes (\cref{fig:A2}). On the opposite side, fainter sources are more prone to contamination by non-AGN populations due to increasing photometric uncertainties, an aspect that potentially reduces purity at fainter magnitudes.

\begin{figure*}[]
   \centering
   \includegraphics[width = \textwidth]{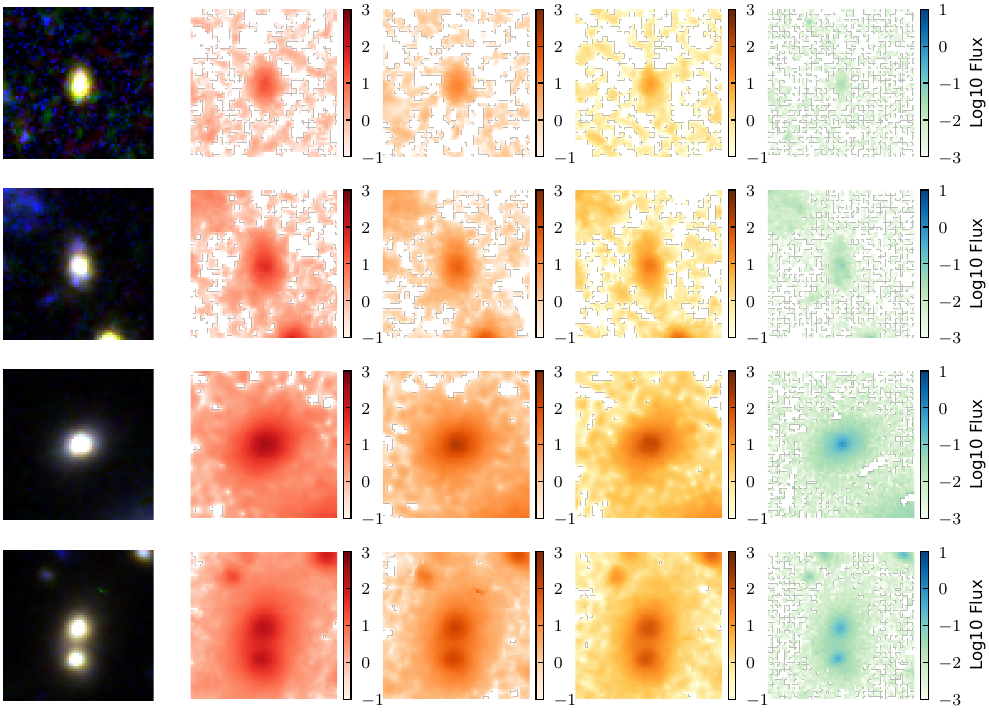}
  \caption{Candidate sources. Panels show, from left to right, the RGB image and the \HE, \JE, \YE and \IE bands. Raw images were obtained through ESA Datalabs \citep{Navarro2024}.}
    \label{fig:ex1}
\end{figure*}

In our analysis, we used a red QSO mock catalogue based on the stacking of SDSS quasar spectra with varying levels of reddening applied. While this approach effectively models the intrinsic quasar properties, it does not account for the host galaxy contamination. The host galaxy flux can dominate over the AGN emission in the optical and near-infrared bands, especially for lower-redshift sources. The consequent dilution of quasar colours may reduce the effectiveness of our HC selection criteria, which are primarily optimised for quasar-dominated SEDs. Such aspects set the seed for future works, where our mock catalogue incorporates realistic galaxy templates, either by adding scaled galaxy SEDs to the QSO spectra or by using simulations informed by empirical measurements of quasar host properties. Nevertheless, our analysis still provides valuable insights into the potential of \Euclid for identifying red quasars. As reported in \cref{sec:results}, we also note that our results remain robust for brighter sources ($H < 22$), aligned closer to SDSS, where the impact of host galaxies is expected to be minimal.

 The probabilistic RF classifier refines the empirical cc-cut by achieving a higher completeness and purity ($98 \%$ and $88 \%$, respectively). This improvement is due to the RF ability to integrate non-linear relationships in the multi-dimensional feature space, which are not detected with simpler colour-colour cuts.

A key aspect in our RF approach is to start from a set of features including the most significant colours, according to PCA, and then expand it with additional colours and magnitudes. Through feature importance analysis, this multi-step methodology allows us to assess how the classification performance is impacted by the information encoded in the additional features. Furthermore, such expansion did not lead to substantial improvements in completeness and purity, confirming the NIR multi-dimensional colour space ($\JE - \HE$, $\YE - \HE$, and $\IE - \HE$) as the most informative and effective to classify out target population.

The UMAP visualisation of the datasets, colour-coded by the probability of being a red quasar, shows that the threshold $P>0.7$ effectively segments the transition region between the two classes. In comparison with the overlayed validation datasets, it shows consistency and robustness against contaminants. 

The visual inspection of the selected sources reveals that most candidates exhibit point-like or slightly extended profiles, consistent with AGN-dominated systems. The multi wavelength images in the first three panels of \cref{fig:ex1} show sources with bright NIR emission and a weaker component in the VIS band, consistent with significant optical light attenuation due to dust. These characteristics suggest that the sources are indeed candidate red QSOs. Also in this case we highlight the importance of this dataset for future morphological analyses, performing host galaxy decomposition and adding information on the properties of the AGN component and its dusty cocoon.

 In addition to examples of individuals red QSOs candidates, in the fourth panel of \cref{fig:ex1} we introduce an example of candidate dual AGN system. The image displays two distinct compact sources with optical-to-NIR colour $I_E - H_E= 2.4$, at redshifts 0.86 (central source) and 0.9, classified as candidate red quasars with RF-based probability of 0.82 (central source) and 0.74. Their proximity consists in a projected distance less then $100 \ kpc$ and difference in redshift not surpassing 0.06. Such criteria aligns with the definition of dual AGN given in previous work \citep{2019NewAR..8601525D}. Such systems offer a unique observational window into AGN triggering mechanisms, galaxy mergers, and the evolution of supermassive black holes. 
The example presented in this work will be part of future systematic searches of dual AGN systems. This effort will use morphological analysis and spectroscopic follow-up where available. The results of this investigation will be presented in forthcoming publications.  

We expand this first characterisation of the selected sources making a comparison between candidates selected through VISTA and DECam colours and via \Euclid-only colours. In the EDF-F we identify a sample of 43528 objects observed by \Euclid, the Vista Hemisphere Survey (VHS, \citealt{2019yCat.2359....0M}) and DES. Among them, we select the $3 \%$ and the  $4 \%$ using solely \Euclid and VISTA+DECam colours, respectively. 

\begin{figure}
    \centering
    \includegraphics[width=\linewidth]{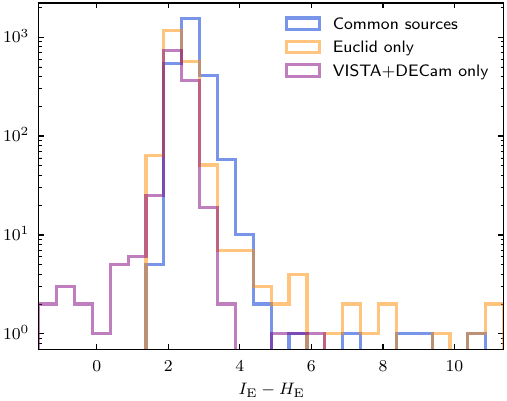}
    \caption{Optical-to-NIR colour distributions for the EDF-F sample of candidate red QSOs observed through both \Euclid and VISTA+DECam photometry. The orange and purple distributions are for candidates selected only via \Euclid and VISTA+DECAm colours, respectively.}
    \label{fig:distributions}
\end{figure}

As shown in \cref{fig:distributions}, \Euclid-only selected objects extend into redder $\IE - \HE$ values. This suggests that \Euclid is better at identifying the reddest sources, which might be missed by VISTA. Furthermore, we observe that the candidates identified trough the VISTA+DECam system display a broader distribution towards bluer colours. This can be explained by VISTA depth and resolution, both inferior to \Euclid, leading to misclassification and missing reddened sources. This preliminary analysis suggests that \Euclid better resolution and NIR sensitivity enables a more complete and robust identification of red QSOs.

\begin{figure*}
   \centering
   \includegraphics[width = \textwidth]{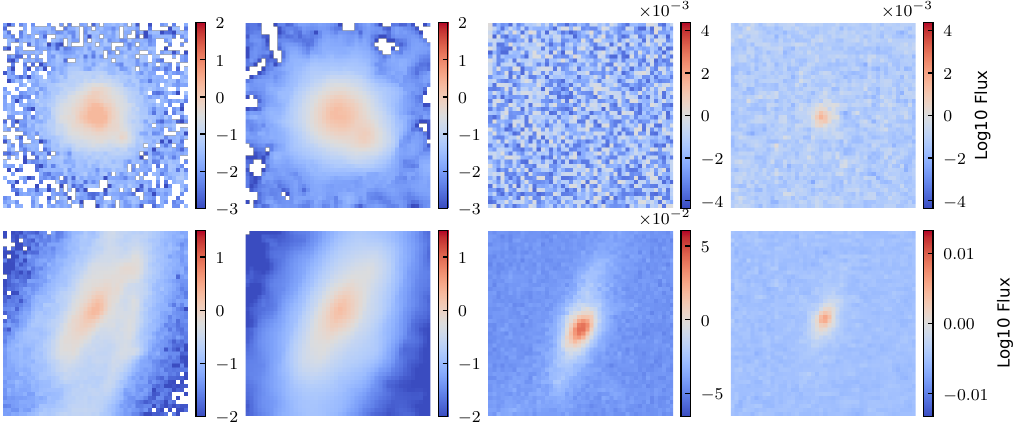}
  \caption{Examples of two candidate red QSOs. The first and the second rows show an \Euclid only and VISTA+DECam-only selected candidate, respectively. Panels from left to right display the object in the \IE and \HE filters, in DECAm \textit{i} and VISTA \textit{$K_s$}.}
    \label{fig:cross_selection}
\end{figure*}

The first panel of \cref{fig:cross_selection} shows an example of red QSO candidate selected with \Euclid only. The image in the $\IE$ band shows a compact source with a bright centre and some faint surrounding structure which can be attributed to the quasar host galaxy. The presence of asymmetry in the outer structure hints at a merger history or disturbed morphology. The \Euclid $\HE$ band image shows a smooth compact core with higher emission than in the optical. This proves that the optical-to-NIR contrast is high. The VISTA and DES images are noisier and the source is harder to distinguish. The poorer signal can explain why the VISTA+DECam system did not classify this sources as a candidate red QSO. Images are normalised and in flux units.

We repeat the visual inspection on sources that were selected as candidate red QSOs through the VISTA+DECam system only. In the second panel of \cref{fig:cross_selection}, the \Euclid $\IE$ band displays an extended morphology with clear structures, possibly star-forming regions or satellite companions. In the DECAm $i$ band only the overall structure is less visible due to the lower resolution compared to \Euclid. Similarly, in the VISTA bands the galaxy is less resolved and appears to be more concentrated in NIR. In this case, DECam and VISTA lower resolution makes the system look more compact, while \Euclid sees a more extended morphology with detailed substructural features. Therefore, the VISTA+DECAm colours used for the selection can be biased towards the integrated light of the system rather than its true morphology. This suggest that the \Euclid-selected sample of red QSOs tends to be cleaner, avoiding that extended galaxies are misclassified as quasars. Additional examples are reported in the other panels of the figure.

In this work, we flag over 150\,000 sources in the EDF-F as candidate red QSOs. This population serves as a first base for future refinements, incorporating additional diagnostics such as MIR data, spectral analyses (spectra are not considered in this work) and morphological compactness. These features should help to further improve purity without compromising completeness. In terms of number expectations, we refer to the studies by \citet{2024A&A...691A...1E} and \citet{EP-Selwood}, conducted in anticipation of the Q1 data release. The former predicts a fraction of 57\% AGNs among NIR-selected sources in EDF-F. The latter estimates an obscured fraction of 26\% in the expected AGN population. Applying these forecasts to our EDF-F sample, we calculate 196\,992 expected obscured AGNs. This is compatible with our selected sample, which is set to include reddened AGNs, a contamination of red galaxies, and red QSOs. The classification of these sources will be the subject of future refinements based on the aforementioned criteria.

In this work, we excluded X-ray sources from the COSMOS2020 sample. However, their integration will be beneficial to future analyses, as their properties are directly linked to the degree of obscuration \citep{2024ApJ...974..225M}. This approach, which can be tested against the Q1 catalogue of X-ray AGN counterparts by \citet{Q1-SP003}, can help improve the distinction between truly reddened AGNs from red galaxies, thereby improving the purity of the selected sample.  

\section{Conclusions}

In this work, we explored the capability of selection criteria based on \Euclid optical and NIR photometry to identify and characterize red quasars. We explored the separation in a multi-dimensional colour-colour space, as a function of magnitude and redshift, and we compared it with selections based on VISTA+DECAm photometry. Through a joint PCA and statistical analysis we identified selection functions achieving high completeness and moderate purity. Then, to refine the identification of red QSOs, we used a probabilistic RF classifier, which significantly improved purity ($88 \%$) and maintained high completeness ($98 \%$). This result reflects the capability of this method to identify complex, non-linear relationships in a multi-dimensional colour space, beyond the reach of simple empirical cuts. Additionally, visualization methods such as UMAP confirm a clear separation between red QSOs and contaminants, reinforcing the robustness of our selection criteria.

Our analysis shown that \Euclid-based selection recovers redder quasar candidates than those identified by VISTA+DECam, as evidenced by shifts in the $\IE - \HE$ distribution. This suggests that \Euclid’s superior depth and resolution enable a more complete detection of highly obscured sources, which might be misclassified or missed in lower-resolution surveys. Conversely, we found that some VISTA+DECam-selected sources appear as extended galaxies in Euclid images, indicating that lower-resolution surveys may introduce contamination by compact galaxies in the sample of red QSOs.

A key insight from future morphological inspection of the selected candidates will be to investigate the capability of \Euclid to resolve host galaxy features in a subset of sources, distinguishing between true quasars and compact galaxies.

We used our findings to build a first census of candidate red QSO in \Euclid Q1. The catalogue will be released as a \textit{fits} table to the \textit{Euclid} Collaboration. This work provides the foundation for systematic red QSO searches in the \Euclid Wide Survey (EWS). The proposed selection function will be further refined with additional diagnostics, including mid-infrared data, spectroscopic follow-ups and the analysis of the host morphology. This first census of red QSO candidates in Euclid Q1 represents a significant step towards a more complete understanding of the dusty AGN population and its connection to galaxy evolution.


\begin{acknowledgements}
\AckQone \AckEC \AckCosmoHub 
This research makes use of ESA Datalabs (datalabs.esa.int), an initiative by
ESA’s Data Science and Archives Division in the Science and Operations
Department, Directorate of Science.

The VISTA Hemisphere Survey data products served at Astro Data Lab are based on observations collected at the European Organisation for Astronomical Research in the Southern Hemisphere under ESO programme 179.A-2010, and/or data products created thereof.

This work has benefited from the support of Royal Society Research Grant RGS{\textbackslash}R1\textbackslash231450.

This research was supported by the International Space Science Institute (ISSI) in Bern, through ISSI International Team project \#23- 573 “Active Galactic Nuclei in Next Generation Surveys”.

F. R., B. L. acknowledge the support from the INAF Large Grant “AGN \& Euclid: a close entanglement” Ob. Fu. 01.05.23.01.14.

\end{acknowledgements}

%
%

\bibliography{Bibliography, Q1, Euclid}

%
%

\begin{appendix}
\clearpage 
\newcommand{\refer}[1]{,\\ \cite{#1}; [{\tt #1}]}
\newcommand{\itemm}{

\medskip\noindent}

\section{Supplementary information\label{apdx:A}}
\begin{table}[h!]
\centering
\caption{The characteristics of filters in COSMOS2020 used for template fitting. The central wavelength correspond to the filter mean wavelength weighted by transmission. Filter names in bold refer to the reconstructed photometry.}
\begin{tabularx}{\columnwidth}{Xccc}
\toprule
\textbf{Filter Name} & \textbf{Central} $\lambda$ ($\text{\AA}$) & \textbf{Bandwidth ($\text{\AA}$)} \\
\midrule
$\mathrm{MegaCam \ CFHT \ \it u}$ & 3682 & 598  \\
$\mathrm{SuprimeCam \ IA427}$ & 4263 & 207  \\
$\mathrm{SuprimeCam \ \it B}$ & 4454 & 892  \\
$\mathrm{SuprimeCam \ IA464}$ & 4635 & 218  \\
$\mathrm{SuprimeCam \ \it g}$ & 4771 & 1265  \\
$\mathrm{HSC \ \it g}$ & 4812 & 1500  \\
$\mathrm{\textbf{DECam \ \it g}}$ & 4826 & 1480  \\
$\mathrm{SuprimeCam \ IA484}$ & 4849 & 229  \\
$\mathrm{SuprimeCam \ IA505}$ & 5062 & 231  \\
$\mathrm{SuprimeCam \ IA527}$ & 5261 & 243  \\
$\mathrm{SuprimeCam \ \it V}$ & 5464 & 1900  \\
$\mathrm{SuprimeCam \ IA574}$ & 5764 & 273  \\
$\mathrm{HSC \ \it r}$ & 6230 & 1547  \\
$\mathrm{SuprimeCam \ IA624}$ & 6232 & 300  \\
$\mathrm{SuprimeCam \ \it r}$ & 6274 & 1960  \\
$\mathrm{\textbf{DECam \ \it r}}$ & 6432 & 1480  \\
$\mathrm{SuprimeCam \ IA679}$ & 6780 & 336  \\
$\mathrm{SuprimeCam \ IA709}$ & 7075 & 316  \\
$\mathrm{\textbf{Euclid \ VIS \IE}}$ & 7180 & 3900  \\
$\mathrm{SuprimeCam \ IA738}$ & 7360 & 324  \\
$\mathrm{SuprimeCam \ \it i}$ & 7667 & 2590  \\
$\mathrm{SuprimeCam \ IA767}$ & 7686 & 365  \\
$\mathrm{HSC \ \it i}$ & 7702 & 1471 \\
$\mathrm{\textbf{DECam \ \it i}}$ & 7826 & 1470 \\
$\mathrm{SuprimeCam \ IA827}$ & 8244 & 343 \\
$\mathrm{HSC \ \it z}$ & 8903 & 766 \\
$\mathrm{SuprimeCam \ \it z^{+}}$ & 9041 & 847 \\
$\mathrm{SuprimeCam \ \it z^{++}}$ & 9099 & 1335 \\
$\mathrm{\textbf{DECam \ \it z}}$ & 9178 & 1520 \\
$\mathrm{HSC \ Y}$ & 9771 & 1810 \\
$\mathrm{UltraVISTA \ \it Y}$ & 10214 & 923 \\
$\mathrm{\textbf{Euclid \ NISP \ \YE}}$ & 10858 & 2630 \\
$\mathrm{UltraVISTA \ \it J }$ & 12535 & 1718 \\
$\mathrm{\textbf{Euclid \ NISP \ \JE}}$ & 13685 & 4510 \\
$\mathrm{UltraVISTA \ \it H}$ & 16454 & 2905 \\
$\mathrm{\textbf{Euclid \ NISP \ \HE}}$ & 17739 & 5670 \\
$\mathrm{UltraVISTA \ \it K_s}$ & 21540 & 3074 \\
$\mathrm{Spitzer \ IRAC \ \it I1}$ & 35313 & 7443 \\
$\mathrm{Spitzer \ IRAC \ \it I2}$ & 44690 & 10119 \\
\bottomrule
\end{tabularx}
\label{tab:filters}
\end{table}

\begin{table}[h!]
\centering
\caption{Principal Component Coefficients for each VISTA HC.}
\begin{tabular}{lrrr}
\hline
\textbf{Feature} & \textbf{HC1} & \textbf{HC2} & \textbf{HC3} \\
\hline
$J-H$   &  0.490950 & $-$0.673444 & $-$0.297702 \\
$H-K_s$ &  0.485117 &    0.738134 & $-$0.267680 \\
$J-K_s$ &  0.521204 & $-$0.036960 & $-$0.303254 \\
$i-K_s$ &  0.501972 & $-$0.016317 &   0.864730 \\
\hline
\end{tabular}
\label{tab:pca_coefficients_uvista}
\end{table}

\begin{table}[h!]
\centering
\caption{Principal Component Coefficients for each \Euclid-like HC.}
\begin{tabular}{lrrr}
\hline
\textbf{Feature} & \textbf{HC1} & \textbf{HC2} & \textbf{HC3} \\
\hline
$Y-J$ &  0.494312 & $-$0.675909 & $-$0.345271 \\
$J-H$ &  0.492558 &    0.732726 & $-$0.220114 \\
$Y-H$ &  0.513889 &    0.021720 & $-$0.295104 \\
$I-H$ &  0.498959 & $-$0.076081 &    0.863280 \\
\hline
\end{tabular}
\label{tab:pca_coefficients_euclid}
\end{table}

\begin{figure}[h!]
   \centering
   \includegraphics[width=\columnwidth]{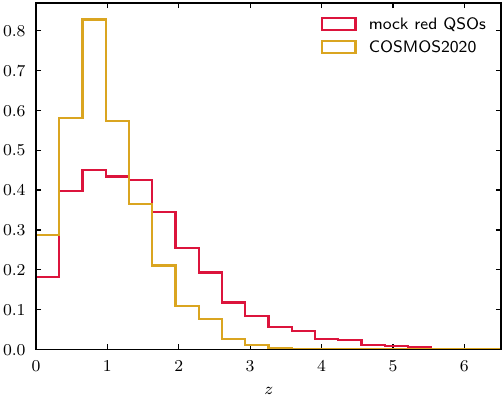}
   \caption{Redshift distributions of the mock and the COSMOS2020 dataset. For the latter, we consider the photometric redshift calculated with \texttt{LePhare} and available in The Classic catalogue.}
    \label{fig:A1}
\end{figure}

\begin{figure}[h!]
   \centering
   \includegraphics[width=\columnwidth]{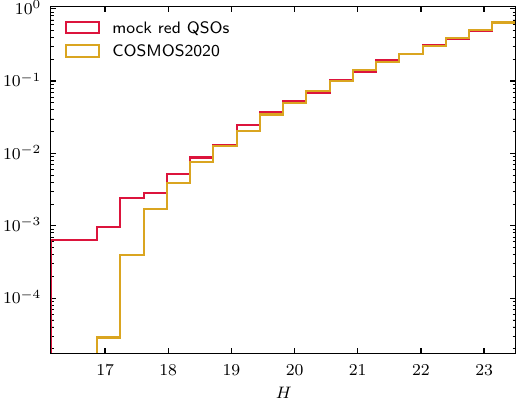}
   \caption{VISTA $H$ magnitude distributions of the mock and the COSMOS2020 dataset.}
    \label{fig:A2}
\end{figure}

\end{appendix}

\label{LastPage}
\end{document}